%% file: zero_temp-arxiv.tex
\documentclass[prb,singlecolumn,a4paper,amsmath,amssymb,superscriptaddress,nofootinbib]{revtex4-2}

\usepackage{graphicx}
\usepackage{rotating}
\usepackage{amsmath,amssymb}
\usepackage{hyperref}
\usepackage{xcolor}
\usepackage{pdfpages}
\usepackage{circuitikz}

\makeatletter
\AtBeginDocument{\let\LS@rot\@undefined}
\makeatother

\newcommand{\Fref}[1]{Fig.~\ref{#1}}
\newcommand{\Eqref}[1]{Eq.~(\ref{#1})}
\newcommand{\nn}{\nonumber}
\newcommand{\be}{\begin{equation}}
\newcommand{\ee}{\end{equation}}
\newcommand{\bear}{\begin{eqnarray}}
\newcommand{\eear}{\end{eqnarray}}
\newcommand{\im}{\mathrm{i}}
\newcommand{\jm}{\mathrm{j}}

\newcommand{\kb}{k_\mathrm{_B}}

\newcommand{\md}{\mathrm{d}}

\begin{document}

\title[Set-up for study of thermal noise, absolute temperature and Boltzmann constant]
{Set-up for observation thermal voltage noise and determination of
absolute temperature and Boltzmann constant}

\author{Todor~M~Mishonov}
\email[E-mail: ]{mishon@bgphysics.eu}
\affiliation{Georgi Nadjakov Institute of Solid State Physics, Bulgarian Academy of Sciences,
72 Tzarigradsko Chaussee Blvd., BG-1784 Sofia, Bulgaria}

\author{Nikola~S~Serafimov}
\affiliation{ Institute of Nuclear Physics and Nuclear Energy, Bulgarian Academy of Sciences,
72 Tzarigradsko Chaussee Blvd., BG-1784 Sofia, Bulgaria}

\author{Emil~G~Petkov}
\affiliation{Union of Physicists in Bulgaria, 5 James Bourchier Blvd., BG-1164 Sofia, Bulgaria}

\author{Albert~M~Varonov}
\email[E-mail: ]{varon@bgphysics.eu}
\affiliation{Georgi Nadjakov Institute of Solid State Physics, Bulgarian Academy of Sciences,
72 Tzarigradsko Chaussee Blvd., BG-1784 Sofia, Bulgaria}

\date{19 May, 14:15}

\begin{abstract}
We describe a set-up for measurement of the absolute zero 
by Johnson-Nyquist thermal noise which can be performed within a week in every high-school or university.
Necessary electronic components and technical guidelines
for the construction of this noise thermometer are given.
The operating temperature used is in the tea cup range from ice to boiling water
and in this sense the set-up can be given in the hands of every high school and university physics student.
The measurement requires a standard multi-meter with thermocouple and voltage probe
and gives excellent for education purposes percent accuracy.
The explanation is oriented to university level but
due to the simplicity of the explanation motivated 
high-school students can follow the explanation
derivation of the used formulas for determination of
the absolute zero and the Boltzmann constant.
As a by-product our set-up gives a new method for the 
determination of the spectral density of the
voltage $e_\mathrm{N}$ and current noise $i_\mathrm{N}$ of operational amplifiers.
\end{abstract}

\maketitle

\section{Introduction}

On intuitive level a thermal motion of gas atoms was sensed even in the time of
Democritus and Lucretius Carus~\cite{Jammer:67} but even now not in every university
Maxwell distribution of the molecules is experimentally demonstrated.
The same can be said \textcolor{black}{about} the thermal motion of Brownian particles which requires at least a microscope.
One dimensional motion of electrons along a wire as a tram on 
a rail can be observed by electronic measurements but even in this case the standard equipment of the type of 
TeachSpin~\cite{TeachSpin} is affordable only for the elite faculties of physics.
The purpose of the present article is to describe how a \textcolor{black}{experimental} simple set-up can be made in every university and even high-school within pocket money budget.

\textcolor{black}{One of the purposes of our work is to suggest a new
design of the set-up for measurement of absolute temperature,
Boltzmann constant $\kb$ and electron charge $q_e$ which is hundred times
cheaper than commercially available laboratory equipment.
Hand made set-ups has one additional advantage -- the understanding.}

To whom is addressed this study,
we describe the mixture of levels of the possible readers beginning with the senior one~\cite{epo5:a,epo6:a}:

1. \textcolor{black}{Starting} with ``XL'' category of readers.
they are our colleagues preparing every year
lectures on disciplines Statistical or Thermal physics. 
Our manuscript interpolates between the Einstein~\cite{Einstein:07}, 
Schottky~\cite{Schottky:18}, Johnson~\cite{Johnson:28} and Nyquist~\cite{Nyquist:28} papers and
contemporary textbooks. 
Touching to the scientific archaeology 
following the style of McCombie~\cite{McCombie} 
in the excellent collection by Landsberg
we give an example how the basic physical laws related to electric fluctuations can
be illustrated using contemporary integrated circuits.
\textcolor{black}{This category of readers is typical for Eur. J. Phys.
The theory corresponds to teaching of thermal physics as a theoretical discipline.
Simultaneously, we explain all know-how details corresponding to the typical level
of university teachers preparing new-set ups for experimental work.
The set-up can be used for complimentary demonstration to the lectures on thermal physics
which in many university has different names, for example thermodynamics and statistical physics, or statistical thermodynamics (as chemists call it).
The multiplication of the set-up can be used for laboratory work for the initial years of university education.
}

2. The second ``L'' auditorium consists of young experimenters 
typically at assistant professor position which are designing 
set-ups for the physics laboratories.
\textcolor{black}{
It is not supposed that these readers can follow the complete theoretical derivation
of the used well-known formulae. 
Those formulae are their starting point of understanding how the set-up works.
The detailed description how the set-up can be reproduced is addressed to this auditorium.
We give all technical details of the used electronics in order to avoid additional reading
of the textbooks on the art of electronics.
We carefully cite the specification of the used integral circuits 
and in such a way we build a bridge between the theoretical physics and contemporary electronics; the authors of the present work have overlapping expertise.
Using pocket money budget they can reproduce our set-up in a day and to compare its work
with the same of the more sophisticated set-ups in terms of equipment.
}

3. The third (``M'') level  are university students which are curious to reproduce something
similar on the subject which they study, the mentioned above fundamental constants for example.
There are a lot of examples available on Internet and academic journals can 
improve the standard of the presentation on this direction.

4. Some very motivated high-school students (category ``S'')
which know that temperature is related to 
random thermal motions of the particles.
\textcolor{black}{Having access to a soldering station and an electronics store they can easily
reproduce the set-up and measure 2 fundamental constant at home or at school. 
This low number category actually contains a many our future colleagues.
}

5. Last but not least our 5$^\mathrm{th}$ column are high-school teachers 
which some time ago
were curious and motivated students.
Human curiosity like childhood never passes completely.
We can send by conventional post free of charge a set-up for 
the Physics Teachers which indicate any interest and get in touch with us.
Demonstration of Maxwell distribution of atoms requires
expensive vacuum technique accessible only in \textcolor{black}{high tech} universities
while demonstration of thermal voltage fluctuations becomes available
in the whole world.

Study of electric fluctuations is for decades important 
laboratory work in good universities, for instance Refs.~\cite{Earl:66,Kittel:78,MIT,MIT1,MIT2,SB,Flowers:17,Qu:19,Thaned:18,Thaned:20,TeachSpin}, however all those set-ups are difficult to be reproduced
or require expensive commercial electronics.
The purpose of the present work is to describe and innovation of
teaching of electrical noise with a set-up which can be easily reproduced 
within one week in every school.
Moreover, all formulae of the used electronics are elementary 
and derived as a homework on algebra for undergraduates.

The paper is organized as follows:
after the elementary theory given in the next section \ref{Theory}
we describe our set-up in Sec.~\ref{Set-up} and finally we represent the 
typical experimental data in Sec.~\ref{Experiment}.
In Sec.~\ref{Discussion} we conclude what can be done in order slightly to improve high-school education related to thermal phenomena.

\section{Theory}
\label{Theory}
\subsection{Classical statistics}
We start with recalling the Nyquist theorem~\cite{Nyquist:28} for the spectral density of
the voltage noise $(\mathcal{E}^2)_f$ of an frequency dependent impedance $Z(\omega)$ at temperature $T^\prime$
\begin{align}
(\mathcal{E}^2)_f=4 \mathcal{R}(\omega)\,T,\qquad T=k_\mathrm{B}T^\prime\gg\hbar\omega,\qquad
\mathcal{R}=\Re(\mathcal{Z}),\qquad \omega=2\pi f,
\label{Nyquist}
\end{align}
where $k_\mathrm{B}$ is the Boltzmann constant and $\hbar$ is the Planck one.
The temperature in energy units is denoted by $T$,
while in Kelvins by $T^\prime$,
\textcolor{black}{and $f$ is frequency in Hz}.

As a simple illustration of calculation mean squared voltage using this theorem 
let us analyze parallelly  connected
\textcolor{black}{capacitor with capacitance $C$ and a resistor with resistance $R$}. 
The total admittance is
\begin{align}
\frac1{\mathcal{Z}}=\frac1R+\mathrm{j}\omega C,\qquad\mathrm{j}=-\mathrm{i},
\label{Z_RC}
\end{align}
where $\mathrm{i}$ is the imaginary unit which participate in the formula
for the amplitude of plane electromagnetic waves 
$\exp[\mathrm{i}(\mathbf{k}\cdot\mathbf{r}-\omega t^\prime)]$.
In electronics usually 
$\exp[\mathrm{j}(\omega t^\prime)]=\exp[\mathrm{i}(-\omega t^\prime)]$.
For the real part of the impedance \Eqref{Z_RC} we have 
\begin{align}
\mathcal{R}(\omega)=\frac{R}{1+(\omega\tau)^2},\qquad 
\tau=RC.
\end{align}
The substitution in the Nyquists theorem \Eqref{Nyquist} 
gives for the total mean square of the voltage 
\begin{align}
\langle U^2\rangle=\int_0^\infty (\mathcal{E}^2)_f\, \mathrm{d} f
=\frac{T}C,\qquad
\frac12\, C\langle U^2\rangle=\frac12\,T,
\end{align}
in agreement with the equipartition theorem. 
As the frequency independent white  noise spectral density $4RT$ is the only one which leads to the equipartition theorem where the dissipation coefficient $R$ does not participate
one can conclude that Nyquist theorem is a consequence of the equipartition theorem.
Albert Einstein~\cite{Einstein:07} was the first who suggested 
that Boltzmann constant $\kb$ can be measured using this relation
if the thermally fluctuating voltage on the capacitor can be significantly amplified.
However, our methodological experiment was made quite recently~\cite{epo5}
and the set-up was given to the participant 
of the Experimental Physics Olympiad~\cite{epo5:a}
The drift of the zero was the obstacle this method~\cite{Habicht:10} 
to be used hundred years ago, 
but now the low noise operational amplifiers (OpAmp) allow the construction of a
pre-amplifier with amplification $Y=10^6=10^{60/10}$ 
\textcolor{black}{(million times, 60 decibels in voltage or 120~dB in power)}
and to measure averaged square of the amplified voltage 
~\cite{epo5,epo5:a}
\begin{align}
\langle U_\mathrm{a}^2\rangle=|Y|^2\langle U^2\rangle
=|Y|^2\, \frac{\kb T^\prime}{C},\qquad 
U_\mathrm{a}=Y\,U.
\label{equipartition_C}
\end{align}
Similar university laboratory set-up was used to determine 
electron charge $q_e$~\cite{epo6} and hundred times multiplied set-up
was given at the 6$^\mathrm{th}$ Experimental Physics Olympiad~\cite{epo6:a}.
No doubts this is methodologically instructive but the idea 
obtained contemporary development and
for determination of the absolute temperature $T^\prime$
(i.e. for the new Kelvin) it is quite possible thermal noise
to be accepted in the future as a new standard,
for the development of the metrology, for example, the recent review Ref.~\cite{Qu:19} and references therein.

In order to illustrate the metrological idea to use 
Johnson~\cite{Johnson:28}-Nyquist~\cite{Nyquist:28} noise as a thermometer
let us analyze a simple consequence: the thermal noise of a resistor with resistance
$R_\mathrm{N}$ is amplified $Y$ times and then is applied to a Low-Pass Filter (LPF)
with is actually a voltage divider with transmission coefficient
\textcolor{black}{(transmission function or transmittance)}
\begin{align}
& \Upsilon_\mathrm{LPF}
=\frac{1/\mathrm{j}\omega C_\mathrm{L}}
{R_\mathrm{L}+1/\mathrm{j}\omega C_\mathrm{L}}
=\frac1{1+\mathrm{j}\omega\tau_\mathrm{_L}},\qquad
\left|\Upsilon_\mathrm{LPF}\right|^2
=\frac1{1+(\omega\tau_\mathrm{_L})^2},\\
&
\mathcal{B}_\mathrm{LPF}
=\int_0^\infty\left|\Upsilon_\mathrm{LPF}\right|^2\, \mathrm{d} f=\frac1{4\tau_\mathrm{_L}}
=\frac1{4R_\mathrm{L}C_\mathrm{L}}, \qquad \tau_\mathrm{_L} \equiv R_\mathrm{L}C_\mathrm{L}.
\end{align}
For the filtered voltage $U_\mathrm{f}$ we have in the frequency representation
\begin{align}
U_\mathrm{f}=\Upsilon_\mathrm{LPF}\,U_\mathrm{a},
\qquad 
\Upsilon_\mathrm{LPF}
=\frac{\mathrm{e}^{\jm \varphi_{_\mathrm{L}}}}
{\sqrt{1+(\omega\tau_\mathrm{_L})^2}},
\qquad
\tan \varphi_{_\mathrm{L}}= -\omega\tau_\mathrm{_L}.
\end{align}
and for its mean square
\begin{align}
\langle U_\mathrm{f}^2\rangle
=4R_\mathrm{N}T\,\frac{|Y|^2}{4R_\mathrm{L}C_\mathrm{L}}=4R_\mathrm{N} T B,
\qquad B\equiv \frac{|Y|^2}{4R_\mathrm{L}C_\mathrm{L}}.
\label{U_f^2}
\end{align}
This result differs from the initial one \Eqref{equipartition_C} only with the multiplier
$R_\mathrm{N}/R_\mathrm{L}$ and replacement of $C$ with $C_\mathrm{L}$.

\textcolor{black}{In the general case for arbitrary frequency,
for the dependence of the amplification 
$\Upsilon_\mathrm{a}(\omega)$
of the pre-amplifier 
and the transmission function of the filter $\Upsilon_\mathrm{f}(\omega)$
we have to measure the averaged square of the amplified thermal voltage}
\begin{align}
\langle U_\mathrm{f}^2\rangle
=4R_\mathrm{N}\kb T^\prime \mathcal{B},\qquad
\mathcal{B} \equiv\int_0^\infty\left|\Upsilon_\mathrm{a}\Upsilon_\mathrm{f}\right|^2\,
\mathrm{d}f\approx B
\label{eq:U2}
\end{align}
\textcolor{black}{for the thermal noise thermometry.}
The calculation or measurement
of the so defined pass bandwidth $\mathcal{B}$ is a routine problem of electronic engineering.
For a high-frequency pre-amplifier \textcolor{black}{it is} often necessary to know only 
the bandwidth of the filter 
$\mathcal{B}_\mathrm{LPF}\equiv\int_0^\infty\left|\Upsilon_\mathrm{LPF}\right|^2\, \mathrm{d}f$,
and for the total pass-band width to use the approximation 
of frequency independent amplification $\mathcal{B}\approx Y^2 \mathcal{B}_\mathrm{LPF}.$
Formally for the exact result we have to substitute 
$B\rightarrow\mathcal{B}$ and the bandwidths have dimension frequency.

In any case for the measurement of averaged square of the filtered amplified voltage
$\langle U_\mathrm{f}^2\rangle$ we have to use
analog multipliers for which the output voltage $U_W$ is a product of the two
input voltages $U_X$ and $U_Y$ divided to a specific for the product constant
$U_\mathrm{m}$
\begin{align}
U_W=\frac{U_XU_Y}{U_\mathrm{m}},\qquad U_X=U_Y=U_\mathrm{f}, \qquad
U_\mathrm{f}=\Upsilon_\mathrm{LPF}\,U_\mathrm{a}.
\label{U_W}
\end{align}
These input voltages have to be equal to the filtered amplified by the pre-amplifier voltage $U_\mathrm{f}$.
The technical details are described in the next section \textcolor{black}{but before that next}
we describe in short the low temperature and high-frequency asymptotic.

\subsection{General case of thermal fluctuations}
The low frequency and high temperature approximation
\Eqref{Nyquist}
for the spectral density of the voltage noise is applicable practically for all
practical cases of electronics.
But for some special cases,
imagine mK temperatures and simultaneously GHz frequencies, 
we have to take into account quantum effects.
In this case the temperature in \Eqref{Nyquist} has to be substituted 
by the thermal averaged energy of a quantum oscillator
with inductance $L$ and capacitance $C$
\begin{align}
T\rightarrow E_\mathrm{osc}
=\left<\frac12 L \hat{I}^2+\frac12  \frac{\hat{Q}^2}{C}\right>
=\frac{\hbar\omega_0/2}{\tanh(\hbar\omega_0/2T)},\qquad
L\hat{I}\,\hat {Q}-\hat {Q}\,L\hat{I}=-\im\hbar, 
\qquad \omega_0\equiv\frac1{\sqrt{LC}},
\label{high_frequencies}
\end{align}
\textcolor{black}{where $\hat{~}$ denotes an operator,
$\hat{Q}$ is the quantum operator of the charge and
$L\hat{I}$ is the same of the canonical variable analogous to the momentum}.
For the proof it is necessary to apply the detailed balance principle
to the quantum oscillator exchanging energy with a resistor
with resistance $R$ immersed in environment 
with temperature $T^\prime$.
And we arrive at the general result
\begin{align}
(\mathcal{E}^2)_{\! f}=4 \mathcal{R}(\omega_0)\,
\frac{\hbar\omega_0/2}{\tanh(\hbar\omega_0/2T)}.
\label{Quantum_Nyquist}
\end{align}
The proof is very simple~\cite{Reggiani:19,Reggiani:20,FDT}, for the total impedance of the sequential resonance circuit we have 
\be
\mathcal{Z}=R+\jm\omega L+\frac1{\jm\omega C},\qquad
|\mathcal{Z}|^2=R^2+\left(\omega L-\frac1{\jm\omega C}\right)^2.
\ee
Then for the spectral densities for the current and charge on the capacitor we have
\begin{align}
(I^2)_{\! f}=\frac{(\mathcal{E}^2)_{\! f}}{|\mathcal{Z}|^2},\qquad
(Q^2)_{\! f}=\omega^2(I^2)_{\! f}.
\end{align}
And finally elementary integration of the spectral densities 
of the energy gives
\textcolor{black}{the averaged energy of the oscillator \Eqref{high_frequencies}}
\begin{align}
E_\mathrm{osc}=\int_0^\infty
\left(\frac12L(I^2)_{\! f}+\frac1{2C}(Q^2)_{\! f}\right)\md f
=\hbar\omega_0\left(\frac1{\exp(\hbar\omega_0/T)-1}+\frac12\right)
\end{align}
and in \Eqref{Quantum_Nyquist} the auxiliary index ``$_0$'' 
from \Eqref{Quantum_Nyquist} can be omitted.
The general formula for the spectral density of the voltage can be derived 
using the Einstein approach of the detailed balance principle applied
to a quantum oscillator.
Further applying casual principle one can derive the fluctuation-dissipation theorem
as a consequence of Nyquist theorem~\cite{FDT}.

\section{Experimental Set-up}
\label{Set-up}

\begin{figure}[ht]
\centering
\includegraphics[scale=0.4]{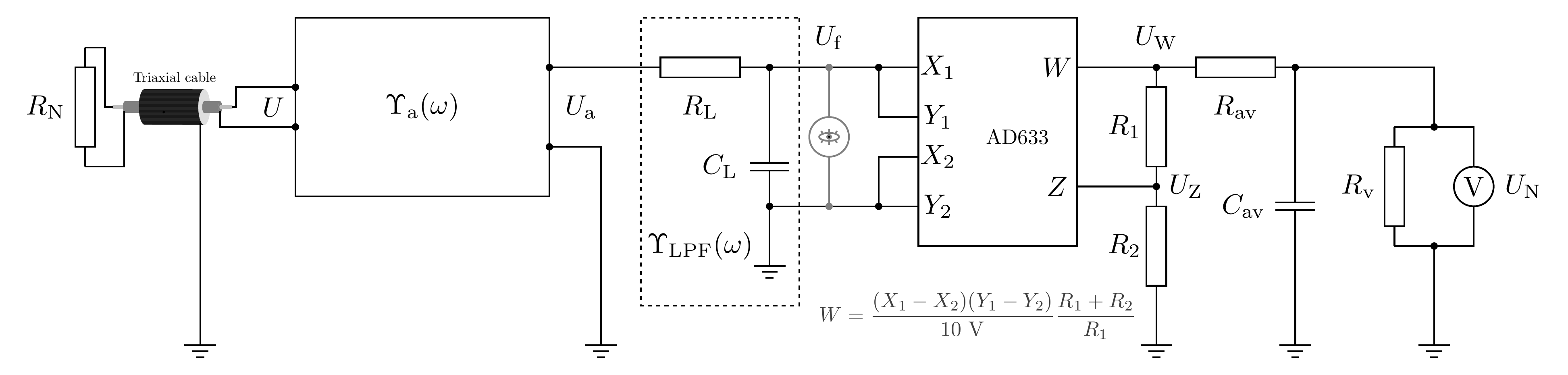}
\caption{Schematic representation of the experimental set-up,
consisting from left to right of a noise resistor $R_N$ whose thermal voltage fluctuations
\textcolor{black}{$U$} are to be measured,
a triaxial cable \textcolor{black}{which inner conductor and shield are connected to $R_N$.
The outer shield} is grounded to the common point of the pre-amplifier;
this is crucial know-how to avoid ringing.
Actually the use of a triaxial cable is crucial innovation of the described set-up,
without it is \textcolor{black}{extremely difficult} to make noise thermometry and cooling and heating of the resistor is significantly more difficult.
The pre-amplifier with $\Upsilon_\mathrm{a}(\omega)$ 
amplification shown in detail in \Fref{Fig:PreAmp}, a low-pass filter with
$\Upsilon_\mathrm{LPF}(\omega)$ response, 
an AD633 analog multiplier with given formula for 
multiplication~\cite[Fig.~17]{AD633}, an averaging filter and a voltmeter.
A repeater may be added before to the voltmeter,
or else the internal resistance of the voltmeter $R_\mathrm{V}$
has to be taken into account in this final voltage divider.
\textcolor{black}{
For visual illustration of the amplified thermal noise
an oscilloscope marked with an eye is necessary to be connected.
Tracing the path of the noise signal:
the noise resistor $R_N$ creates a thermal noise voltage $U$
which is amplified by the pre-amplifier to $U_\mathrm{a}$
and then filtered by a Low Pass Filter (LPF) to $U_\mathrm{f}$.
This filtered voltage $U_\mathrm{f}$ is further applied to the 2 inputs 
$X_1$ and $Y_1$ of a multiplier as shown in
Eq.~\ref{amplified_filtered_applied}.
The output voltage of the multiplier $W$
is divided by the voltage divider with resistors $R_1$ and $R_2$.
The divided voltage $U_Z$ is applied to the adding input of the multiplier;
in \Eqref{multiplier_amplification} we use the simplified notations $W\equiv U_W$
and analogously $Z\equiv U_Z$.
Finally the voltmeter shows the averaged output voltage of the multiplier
passed trough a voltage divider 
$U_N=\langle W \rangle R_\mathrm{V}/(R_\mathrm{V}+R_\mathrm{av})$, 
see \Eqref{W} and \Eqref{averaged square}.
If additionally a shot noise current from a photo-diode is applied through $R_N$,
as it is described in great detail in Refs.~\cite{epo6,epo6:a} this set-up can also be used for determination of the electron charge $q_e$.
}
}
\label{Fig:Schema}
\end{figure}

\begin{figure}[ht]
\centering
\includegraphics[scale=0.4]{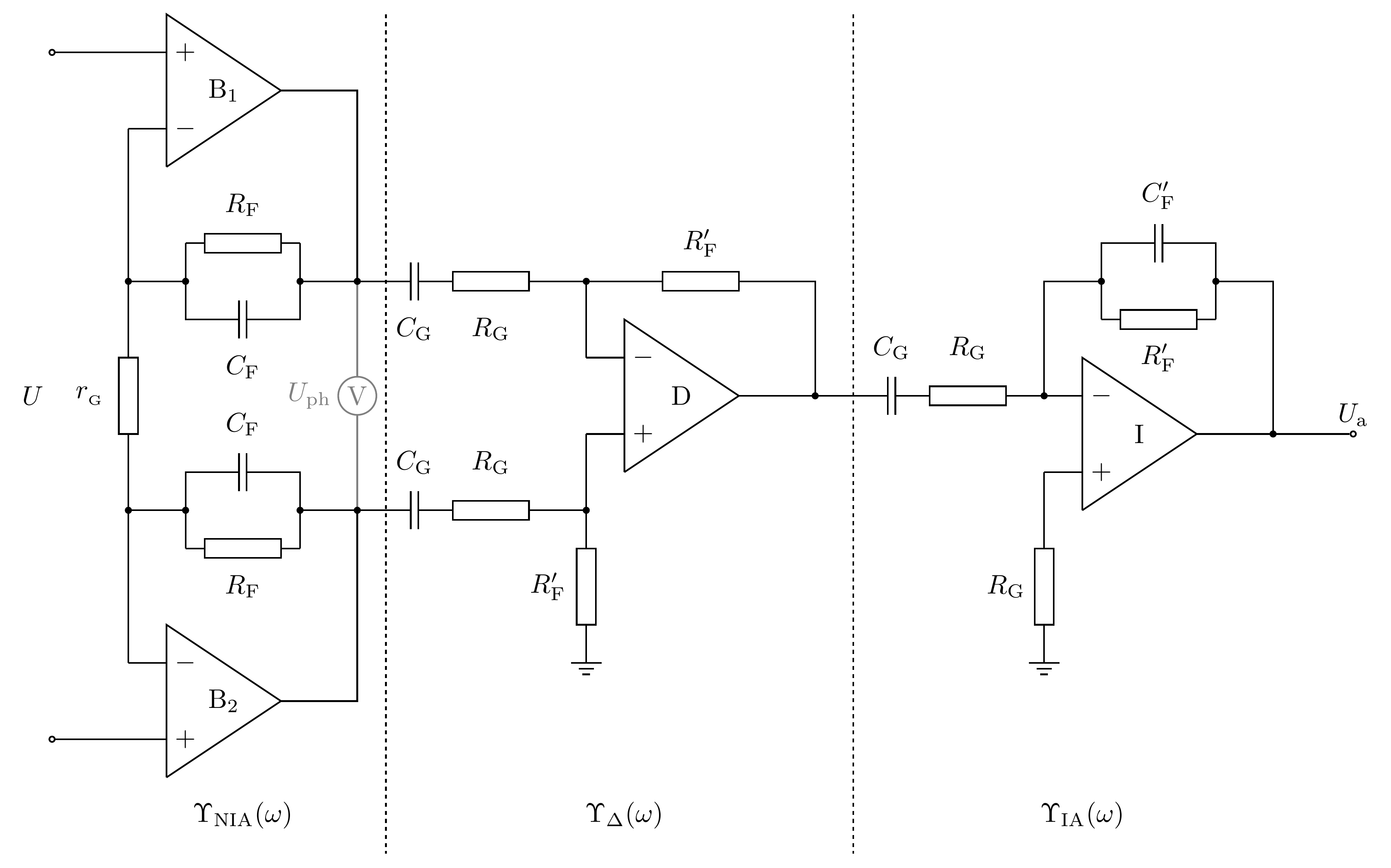}
\caption{Circuit representation of the used pre-amplifier in the experimental set-up schematically presented in \Fref{Fig:Schema}.
The pre-amplifier consists of coupled NIA or buffer with amplification $\Upsilon_\mathrm{NIA}(\omega)$ to the left of the first dashed vertical,
followed by a difference amplifier in the middle between the two verticals
with amplification $\Upsilon_\Delta(\omega)$
and ending with an IA to the right with amplification $\Upsilon_\mathrm{IA}(\omega)$.
The whole amplification of the pre-amplifier
$\Upsilon_\mathrm{a}(\omega) = \Upsilon_\mathrm{NIA}(\omega) \Upsilon_\Delta(\omega) \Upsilon_\mathrm{IA}(\omega)$.
The first couple ADA4898-2~\cite{ADA4898} OpAmps labeled B$_1$ and B$_2$ are for the buffer or coupled NIA amplifiers, while the second couple are for the difference amplifier with label D and for the IA with label I. 
The first two amplifiers $\Upsilon_\mathrm{NIA}(\omega) \Upsilon_\Delta(\omega)$
are actually one instrumentation amplifier. 
The voltage supplies for the OpAmps are omitted here for brevity and given in \Fref{Fig:Supply}.
}
\label{Fig:PreAmp}
\end{figure}

In our set-up drawn schematically in \Fref{Fig:Schema} 
a multiplier AD633~\cite{AD633} for which $U_\mathrm{m}=10\,$V.
It is self-understanding that drawing the schematics 
the resistance of wires and paths in the printable circuit boards 
is much smaller than the resistance of the resistors 
and the capacity of the triaxial cable is much smaller than the capacity of the
low-pass filter $C_\mathrm{L}$.
Two of voltage inputs are grounded $X_2=Y_2=0$
and the filtered amplified voltage 
\begin{align}
X_1=Y_1=U_\mathrm{f},\qquad U_\mathrm{f}=\Upsilon_\mathrm{LPF}\,U_\mathrm{a}
\label{amplified_filtered_applied}
\end{align}
is applied to the other two.
\textcolor{black}{During measurement $|U_\mathrm{a}|<1\,\mathrm{V}$ to avoid non-linear effects.}
Let us recall the general formula for this multiplier~\cite[Eq.~1, Fig.~12 and Fig.~17]{AD633}
\be
W=\frac{(X_1-X_2)(Y_1-Y_2)}{U_\mathrm{m}}+Z, \qquad
Z=\frac{R_2}{R_1+R_2}\,W, \qquad
W=\frac{(X_1-X_2)(Y_1-Y_2)}{U_\mathrm{m}}\,\frac{R_1+R_2}{R_1}.
\label{multiplier_amplification}
\ee
Elementary substitution of these formulae from the specification gives
\begin{align}
W=\frac{U_\mathrm{f}^2}{\tilde{U}},\qquad 
\tilde{U}=\frac{R_1}{R_1+R_2}\,U_\mathrm{m}=1\,\mathrm{V},\qquad
R_1=2\,\mathrm{k}\Omega,\qquad
R_2=18\,\mathrm{k}\Omega.
\label{W}
\end{align}
The voltage W directly proportional to the square of the filtered voltage $U_\mathrm{f}$
is applied to another averaging LPF with large time constant
\begin{align}
\tau_\mathrm{av}=R_\mathrm{av}C_\mathrm{av}=15\,\mathrm{s},\qquad R_\mathrm{av}
=1.5\,\mathrm{M}\Omega,\qquad
C_\mathrm{av}=10\,\mu\mathrm{F}. \nn
\end{align}
This averaged voltage is measured by a multimeter with an internal resistance
$R_\mathrm{V}$
of the order of 1 or 10 M$\Omega$ depending on the range of application. 
This DC voltmeter measures the noise created voltage
\be
U_\mathrm{N}
\textcolor{black}{\,=
\frac{R_\mathrm{V}\,\langle W \rangle}{R_\mathrm{V}+R_\mathrm{av}}\,}
\label{averaged square}
=
\frac{R_\mathrm{V}}{R_\mathrm{V}+R_\mathrm{av}}\,
\frac{\langle U_\mathrm{f}^2\rangle}{\tilde{U}}
=\frac{\langle U_\mathrm{f}^2\rangle}{U^*},\qquad
U^*\equiv\frac{R_\mathrm{V}+R_\mathrm{av}}{R_\mathrm{V}}\,\tilde{U}
=\frac{R_\mathrm{V}+R_\mathrm{av}}{R_\mathrm{V}}
\frac{R_1}{R_1+R_2}\,U_\mathrm{m}.
\ee
Here we take into account another voltage divider.
After substitution of $\langle U_\mathrm{f}^2\rangle$ from \Eqref{U_f^2} 
\begin{align}
U_\mathrm{N}
=\frac{B}{U^*}\,
4R_\mathrm{N}\,\kb T^\prime,
\end{align}
\textcolor{black}{where $\kb T^\prime$}
is the interesting for us spectral density of the thermal voltage
but it is not the whole truth.
To this spectral density we have to add the voltage noise
of the first operational amplifier $e_\mathrm{N}\approx 0.9\,\mathrm{nV/\sqrt{Hz}}$~[Table~1]\cite{ADA4898}
and the current noise  $i_\mathrm{N}\approx 2.4\,\mathrm{pA/\sqrt{Hz}}$.
In such a way the final measurable voltage reads
\be
U_\mathrm{N}=\left[e_\mathrm{N}^2+ 4R_\mathrm{N}\kb T^\prime
+R_\mathrm{N}^2i_\mathrm{N}^2\right]/\mathcal{A},\qquad
1/\mathcal{A}\equiv\frac{\mathcal{B}}{U^*}=
\frac{R_\mathrm{N}}{(R_\mathrm{V}+R_\mathrm{av})}
\frac{(R_1+R_2)}{R_1U_\mathrm{m}}\,\mathcal{B} \,
,\qquad
\mathcal{B}=\frac{|Y|^2}{4R_\mathrm{L}C_\mathrm{L}}
(1-\varepsilon),
\label{pass_band_width}
\ee
where the small correction 
\be
\varepsilon\equiv 1-\frac{\mathcal{B}}{B},
\qquad
\mathcal{B}=(1-\varepsilon)B
\label{eq:err}
\ee
takes into account more precise \textcolor{black}{numerical} integration
for the exact bandwidth \Eqref{eq:U2}
of the pre-amplifier depicted in \Fref{Fig:PreAmp}.
This pre-amplifier is a sequence of a buffer with a double Non Inverting Amplifiers (NIA)
with amplification $\Upsilon_\mathrm{NIA}$
with a virtual ground between them, 
followed by a difference amplifier ($\Delta$) with amplification $\Upsilon_\Delta$
ending with an Inverting Amplifier (IA) with amplification $\Upsilon_\mathrm{IA}$.
In such a way the total amplification $\Upsilon_\mathrm{a}$ is a product of the sequence
\begin{align}
\Upsilon_\mathrm{a}(\omega) 
= \Upsilon_\mathrm{NIA}(\omega) \Upsilon_\Delta(\omega) \Upsilon_\mathrm{IA}(\omega).
\end{align}
The amplification of these amplifiers is given by the well-known formulas
described in many textbooks an specifications;
see, for example, Ref.~\cite[Eq.~4, Eq.~7, Fig.~52 and Fig.~53]{ADA4817}
\begin{align}
&
\Upsilon_\mathrm{NIA}
=\dfrac1{\dfrac1{\dfrac{Z_\mathrm F}{Z_\mathrm G}+1}+\tau s},\qquad
Z_G=\frac{r_{_\mathrm G}}2,\qquad
\frac1{Z_\mathrm F}=\frac1{R_\mathrm F}+s\,C_\mathrm F
,\\
&
\Upsilon_\Delta
=\dfrac1{\dfrac{Z_\mathrm{G}^\prime}{R_\mathrm{F}^\prime}
+\left(\dfrac{Z_\mathrm{G}^\prime}{R_\mathrm{F}^\prime}+1\right)\!\tau s},\qquad
Z_\mathrm{G}^\prime=R_\mathrm{G}+\frac1{sC_\mathrm{G}}
, \qquad \tau=\frac1{2\pi f_\mathrm{c}} \\
&
\Upsilon_\mathrm{IA}
=-\dfrac1{\dfrac{Z_\mathrm{G}^\prime}{Z_\mathrm{F}^\prime}
+\left(\dfrac{Z_\mathrm{G}^\prime}{Z_\mathrm{F}^\prime}+1\right)\!\tau s},\qquad
\frac1{Z_\mathrm{F}^\prime}=\frac1{R_\mathrm{F}^\prime}+sC_\mathrm{F}^\prime,
\end{align}
where $s\equiv\mathrm{j}\omega$ is the purely imaginary at real frequencies $\omega$,
and time-constant  $\tau$
is parameterized by the crossover frequency $f_\mathrm{c}$ of the OpAmps.
For the used dual low noise OpAmp ADA4898-2~\cite{ADA4898}
$f_\mathrm{c}=97.73$~MHz~\cite{PDF}
In short, the initial signal $U$ is  amplified, then filtered, before being squared
\begin{align}
U_\mathrm{f}=\Upsilon_\mathrm{LPF}\,U_\mathrm{a},\qquad
U_\mathrm{a}=\Upsilon_\mathrm{a}\,U
=\Upsilon_\mathrm{IA}\Upsilon_\mathrm{\Delta}\Upsilon_\mathrm{NIA}\,U.
\end{align}
The numerical integration can be easily performed using software
products of the type of Mathematica or Maple,
\textcolor{black}{or even by direct programming with languages such as Python or Fortran.
It is simpler to use the complex $\Upsilon$
and to calculate corresponding modulus $|\Upsilon|^2$.
This calculation of \Eqref{eq:U2} gives $\mathcal{B}=55.44$~\textcolor{black}{P}Hz and from \Eqref{eq:err} $\varepsilon=8.7$\%.}

In the broad frequency interval
\be
\frac1{R_\mathrm{G}C_\mathrm{G}}\ll\omega\ll\frac1{R_\mathrm{F}C_\mathrm{F}}
\ee
the amplification is almost frequency independent
\begin{align}
Y_\mathrm{NIA}=\frac{R_\mathrm{F}}{r_\mathrm{G}/2}+1,
\qquad
Y_\Delta=-Y_\mathrm{IA}=\frac{R_\mathrm{F}^\prime}{R_\mathrm{G}^\prime}, 
\qquad
Y=Y_\mathrm{NIA}Y_\Delta Y_\mathrm{IA}=-1.01\times 10^6.
\label{Y}
\end{align}
The bandwidth of the filter must be in this interval
\be
\frac1{R_\mathrm{G}C_\mathrm{G}}\ll
\frac1{R_\mathrm{L}C_\mathrm{L}}
\ll\frac1{R_\mathrm{F}C_\mathrm{F}}.
\ee

\begin{figure}[ht]
\centering
\includegraphics[scale=0.5]{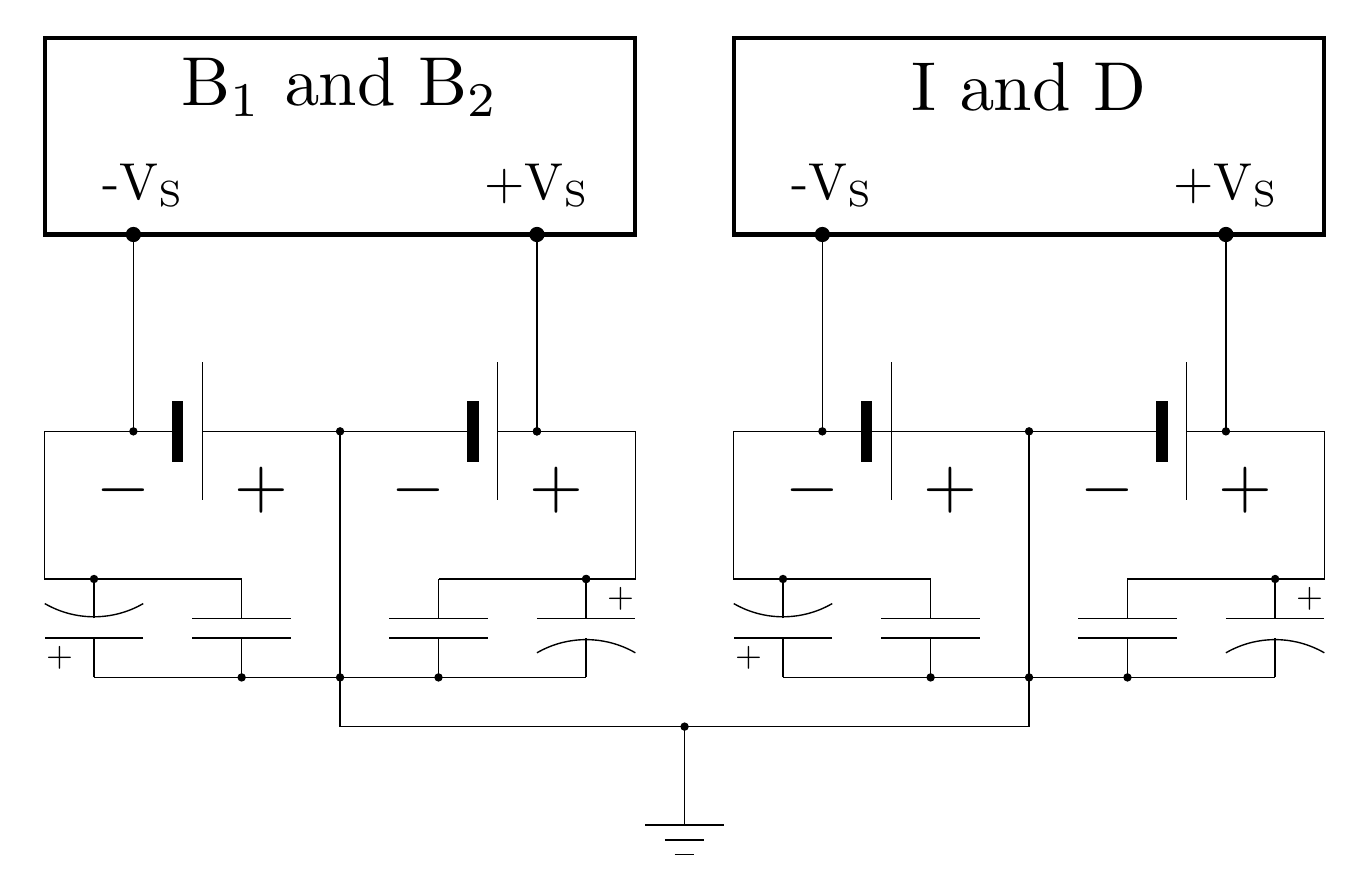}
\caption{Power supplies for the both dual ADA4898-2. 
Every dual OpAmp has different 9~V batteries for its power supply.
Left couple of batteries gives voltage supply $+V_\mathrm{S}=9\,$V
and $-V_\mathrm{S}=9\,$V for the buffer (B$_1$) and (B$_2$), see \Fref{Fig:PreAmp}.
While the right batteries give voltage supply for the OpAmp
of the difference (D) and inverting amplifier (I) from \Fref{Fig:PreAmp}.
In parallel to each of the 2 batteries one electrolytic $10\,\mu$F 
and one ceramic $0.1\,\mu$F capacitor are soldered.
The ground (common point) is connected also to the most external shield 
of the triaxial cable. 
This obligatory screening together with capacitors prevents 
the self-excitation (ringing) of the amplifier with amplification of $10^6$ 
\textcolor{black}{or 60~dB in voltage}. 
The use of two couple batteries makes the set-up much more stable against the ringing
which always appears at big enough noise resistor $R_\mathrm{N}$.
This self excitation can be easily observed by an oscilloscope, but
physical conditions of the stability is an interesting question for further studies
of similar set-ups.
}
\label{Fig:Supply}
\end{figure}

\textcolor{black}{We have also to give technical details related to the power supply of the circuit
shown in \Fref{Fig:Supply}.}
The large 10~$\mu$F capacitors~\cite{WIMA,KEMET} stop the offset voltages of the operational amplifiers ADA4898-2,
their floating of the zero, their low frequency $1/f$~noise and
the low frequency AC voltages coming from the electric power supply in the lab. 
\textcolor{black}{
We use two sets of 9~V batteries in order to guarantee several hours continuous experimental work with the described set-up, it is also possible to use also a couple of 12~V lead batteries.}
The electrolytic capacitors parallel to the batteries ensure damping and 
the fast capacitors  short circuit the undesired feedback which could create self-excitation (ringing) via the voltage supply probes of the operational amplifiers.
These capacitors are suggested in the specification of the used 
ADA4898-2~\cite{ADA4898} and AD633~\cite{AD633} integral circuits.
Finally as the time constant of the final voltage averaging filter is 15 seconds and it is necessary to wait at least 1 minute (4 time constants) in order initial voltage to be ``forgotten'' and measured voltage to be established; $\exp(-4)<2\%.$
\textcolor{black}{This 1~minute does not include the thermalization time of $R_\mathrm{N}$ that is typically several minutes in case the resistor is being heated or cooled.}

We wish to emphasize that the described set-up does not require
specialized commercial electronic equipment like
Butterworth filters, oscilloscopes, computers, preamplifiers and laboratory power supplies
used in many educational noise thermometry experiments~\cite{Earl:66,Kittel:78,Kraftmakher:95,MIT,Thaned:18,Thaned:20};
see also the references therein. 
The electronic circuit is so stable
that it does not even require screening metallic boxes 
and BNC cables between them.
It is pedagogically instructive that all details are visible 
and the only grounding is the connection of the outer shield 
of the triaxial cable to the common point of the pre-amplifier. 
\textcolor{black}{
The used triaxial cable is Multicomp Pro RG403,
practically a coaxial cable with a second additional shield.}
In such a way the small noise signal is transmitted \textcolor{black}{reliably} to the inner
coaxial cable and the parasitic capacitive feed between the amplified
signal and input probes of the buffer is reliably minimized
\textcolor{black}{and the ringing is avoided.}

\textcolor{black}{Summarizing, all values of the circuit elements of the set-up are given in Table~\ref{tbl:values}.}
\begin{table}[h]
\centering
	\caption{Table of the numerical values of the circuit elements from \Fref{Fig:Schema} and \Fref{Fig:PreAmp}. }
\begin{tabular}{ c  r }
		\hline
		Circuit element or parameter & Value  \\ \hline
		     $R_\mathrm{N}$ & 100~$\Omega$ \\
			$R_\mathrm{L}$ & 100~$\Omega$ \\
			$C_\mathrm{L}$ & 42~nF \\
			$1/2\pi R_\mathrm{L}C_\mathrm{L}$ & 37.9~kHz\\
			$C_\mathrm{cable}$ & 96.5~pF/m \\
			$r_\mathrm{cable}$ & 0.8~$\Omega$/m\\
			$r_\mathrm{_G}$ & 20~$\Omega$ \\
			$R_\mathrm{F}$ &  1~k$\Omega$  \\
			$C_\mathrm{F}$ &  10~pF  \\ 
			$C_\mathrm{G}$ & 10~$\mu$F \\
			$R_\mathrm{G}$ &  100~$\Omega$  \\ 
			$ 1/2\pi R_\mathrm{G}C_\mathrm{G}$ & 159~Hz\\
			$R_\mathrm{F}^\prime$ & 10~k$\Omega$ \\
			$C_\mathrm{F}^\prime$ & 10~pF \\
			$R_1$ &  2~k$\Omega$  \\ 
			$R_2$ & 18~k$\Omega$  \\
			$U_0$ & 10~V\\
			$R_\mathrm{av}$ & 1.5~M$\Omega$ \\
			$C_\mathrm{av}$ & 10~$\mu$F \\
			$R_\mathrm{av}C_\mathrm{av}$ & 15~s\\
			$R_\mathrm{V}$ & 1~M$\Omega$ \\
			$+V_\mathrm{S}$ & +9~V~\cite{ADA4898,AD633} \\
			$-V_\mathrm{S}$ & -9~V~\cite{ADA4898,AD633} \\
			$e_\mathrm{N}$ & 0.9~nV/$\sqrt{\mathrm{Hz}}$~\cite[Table~1]{ADA4898}\\
			$i_\mathrm{N}$ & 2.4~pA/$\sqrt{\mathrm{Hz}}$~\cite[Table~1]{ADA4898}\\
			\hline
\end{tabular}
	\label{tbl:values}
\end{table}


Before measuring electric fluctuations, it is necessary to determine the ``eigen''-noise 
of the amplifier,
terminating the triaxial cable with a short circuit, i.e. $R_\mathrm{N}=0$.
Rewriting the final result \Eqref{pass_band_width}
for experimental data processing is convenient to use the variable 
and its polynomial regressions
\begin{align}
\mathcal{S}\equiv
\mathcal{A}U_\mathrm{N}=
e_\mathrm{N}^2+ 4R_\mathrm{N}\kb T^\prime
+R_\mathrm{N}^2i_\mathrm{N}^2,\qquad
\mathcal{A}\equiv\frac{U^*}{\mathcal{B}}
=\frac{(R_\mathrm{V}+R_\mathrm{av})\,R_1\,U_\mathrm{m}}
{R_\mathrm{V}\,(R_1+R_2)\,\mathcal{B}}=45\,\mathrm{aV/Hz}
\label{V^2/Hz}.
\end{align}
The variable $\mathcal{S}$ has dimension V$^2$/Hz;
this is actually the total low frequency spectral density of the voltage noise.
The noise-meter calibration constant $\mathcal{A}$
is the most important parameter of the device giving
the connection between the measured voltage $U_\mathrm{N}$
and the spectral density of the voltage noise $\mathcal{S}$.
The quantity $q_e\mathcal{A}$ has dimension of action 
(energy times time as Planck constant $\hbar$).
Our set-up is a device for measurement of the low frequency spectral density of the voltage noise.
The constant voltage measured by a DC voltmeter $U_\mathrm{N}$
is proportional to the low frequency 
spectral density of the noise $\mathcal{S}$, 
and $\mathcal{A}$ is a constant specific for the noise-meter 
which can be expressed by the parameters of the circuit.

Here we have to make a special clarification.
Administrators related to education could say that spectral density
is a sophisticated notion belonging to final courses of university education.
However, spectral density in sense of energy per unit
frequency is used in every high-school textbook in which black body radiation is described.
Omitting units of the ordinate intensity of black-body radiation
as function of frequency or wavelength is drawn in textbooks for
a century.
Let us introduce dimensionless frequency
\begin{align}
w=\frac{\hbar\omega}{T}=\frac{h f}{\kb T^\prime},\qquad
\omega=2\pi f,\quad
h=2\pi \hbar,\quad
T=\kb T^\prime.
\end{align}
In space dimension $D$
the intensity of the black-body radiation is proportional to the function
\begin{align}
I(\omega;\,D)\propto \frac{w^{D}}{\exp(w)-1};
\end{align}
theoretical explanation and experimental observation of this universal spectral density in $D=3$ has been awarded by 4 Nobel prizes
(Planck 1918, Penzias \& Wilson 1978, Mather \& Smooth 2006, Peebles 2019).
The original consideration of thermal voltage fluctuations 
\Eqref{Nyquist}
by Nyquist~\cite{Nyquist:28} is based on the 1 dimensional
black-body radiation and its low frequency limit
\begin{align}
I(\omega\ll 1;\,D=1)\propto \frac{w}{\exp(w)-1}\approx 1.
\end{align}
This low frequency spectral density $\mathcal{S}$ 
is measured by our set-up.

For short circuit in the input (front-end)
we obtain a new method for determination of the voltage noise of the OpAmp
\begin{align}
e_\mathrm{N}=\sqrt{\mathcal{S}_0},\qquad 
\mathcal{S}_0\equiv\mathcal{S}(R_\mathrm{N}=0).
\end{align}
At fixed temperature $T^\prime$ we can perform measurement of $\mathcal{S}_i$
for a set of noise resistors $R_\mathrm{N},i$ for $i=1,2,\dots,N$.
The parabolic fit of the data $\mathcal{S}$ versus $R_\mathrm{N}$
\begin{align}
\mathcal{S}=a_0+a_1R_\mathrm{N}+a_2R_\mathrm{N}^2
\label{spectral_density}
\end{align}
gives a by product a new method for determination of the current noise and Boltzmann constant
\begin{align}
e_\mathrm{N}=\sqrt{\mathcal{S}_0}= \sqrt{a_0},\qquad
a_1=\left.\frac{\md\mathcal{S}}{\md R_\mathrm{N}}
\right|_{R_\mathrm{N}=0},\qquad
\kb =\frac{a_1}{4T^\prime},\qquad
a_2=\left.\frac{\md^2\mathcal{S}}{2\,\md R_\mathrm{N}^2}
\right|_{R_\mathrm{N}=0},\qquad
i_\mathrm{N}=\sqrt{a_2}.
\label{Boltzmann_constant}
\end{align}
The derivatives here are related to the parabolic fit of the dependence
$\mathcal{S}(R_\mathrm{N})$.
Here we hypocritically suppose that we know the absolute temperature.
However, studying thermal noise it is most interesting to perform
temperature measurements and to try to determine the absolute zero.
If we use small noise resistor $R_\mathrm{N}=50\,\Omega$
the quadratic term $\propto a_2$ has negligible influence.
For a set of temperatures $t_j^\prime$ (measured in Celsius degrees) 
we can measure $\mathcal{S}_j$ and additionally the of-sets $\mathcal{S}_{0,j}$ for short circuits $R_\mathrm{N}=0\,\Omega$.
Then we can perform the linear regression and to extrapolate the 
absolute zero temperature $t_0^\prime$
\begin{align}
\mathcal{S}-\mathcal{S}_0
=b_0+b_1 t^\prime=(t^\prime-t_0^\prime)\,b_1,\qquad t_0^\prime=-b_0/b_1.
\end{align}
At this subtraction cancels the offset of the $U_W$ voltage which actually
has to be written in the right side of \Eqref{U_W}.
\textcolor{black}{The temperature range between ice cold and boiling water is enough for
a reliable measurement meaning that liquid nitrogen or dry ice are not required
but their inclusion would of course dramatically improve the accuracy.}
Operating with tea cup temperatures $t^\prime\in (0,\,80)^\circ$C,
even for a teenager is safe to make the experiment in the kitchen.
The higher temperature can be used is limited by the 
insulation of the triaxial cable 190$^\circ$C. 
But in principle using platinum~100 resistors much higher temperatures can be used in an appropriate laboratory.
For low temperatures however there is no limitation
but the accuracy becomes smaller.
The accuracy is however not essential the purpose is understanding.

Here we have to mention the applicability of the linear regression analysis
since \Eqref{spectral_density} requires parabolic regression with respect to $R_\mathrm{N}$.
In order to be able to analyze linearly this equation,
we need the quadratic term to be negligible
leading to the condition $R_\mathrm{N} \ll e_\mathrm{N}/i_\mathrm{N}=375~\Omega$
according to the parameters from Table~\ref{tbl:values}.
On the other hand according to \Eqref{Nyquist}
the voltage noise of the ADA4898-2 operational amplifier,
 $e_\mathrm{N}=0.9\,$nV
corresponds to the thermal noise of a resistor of 
$e_\mathrm{N}^2/4\kb T_\mathrm{room}^\prime=\,$53.8~K
at $T_\mathrm{room}^\prime=273+25=297\,$K.
The geometric mean $(\sqrt{53.8\times373}=142)\,\Omega$
shows that a noise resistor $R_\mathrm{N}=100\,\Omega$
is close to the optimal value if we wish to study temperature 
dependence if the Johnson noise with our set-up.
As a resistor it is possible to use platinum 100~$\Omega$ resistance temperature detectors
or similar realizations~\cite{MItra:21}.
In this case temperature dependence of the noise resistor
\textcolor{black}{$R_\mathrm{N}(t^\prime)$}
can be used as a good reference thermometer well calibrated in 1~kK interval $t^\prime\in(-200, 800)^\circ$C.
And simultaneously in the same interval we can observe thermal fluctuations of the voltage.
In such a way contemporary technology of low noise operational amplifiers can be
optimally used for educational purposes.
More detailed analysis of the influence of the quadratic term 
comes from the input current noise $i_\mathrm{N}$ is given in
Sec.~\ref{Bilinearity}

\subsection{Additional determination of $\kb$ and $q_e$}

The described device can be used also for determination of the electron charge $q_e$
by Schottky~\cite{Schottky:18,Schottky:22} noise.
In this case to the spectral density of the voltage and current we have to add the contribution by the Schottky shot noise
created by the averaged photo-current
\begin{align}
\mathcal{S}_{+}=R_\mathrm{N}^2i_\mathrm{N,+}^2
=2q_e I_\mathrm{ph}  R_\mathrm{N}^2,\qquad
i_\mathrm{N,+}^2=2q_eI_\mathrm{ph}.
\label{shot_noise}
\end{align}
For a methodological derivation~\cite{epo6}
and application for high-school students see Ref.~\cite{epo6:a}
The electron charge can be determined 
as derivative of the noise spectral density 
$\propto U_\mathrm{N}$ with respect of 
the averaged photo voltage measured after the amplification 
by the buffer
\be
U_\mathrm{ph}=Y_\mathrm{NIA} R_\mathrm{N}I_\mathrm{ph}
\label{photocurrent}
\ee
proportional to averaged photo-current $I_\mathrm{ph}$
(see schematics in \Fref{Fig:PreAmp}).
The story is well-known,
a young post-doc of Planck
(Walter Schottky) attended in a lecture given by Enstein
and impressed by the consideration of fluctuation he suggested 
that electron charge $q_e$ can be also determined by examining of 
current fluctuations.
The electron charge can be expressed from the Schottky formula
for the current noise $i_\mathrm{N,+}^2$ and averaged photo-current
$I_\mathrm{ph}$ \Eqref{shot_noise}.
The excess spectral current density  $i_\mathrm{N,+}^2$
have to be expressed by the 
excess voltage noise density $\mathcal{S}$.
According \Eqref{V^2/Hz}
the voltage spectral density $\mathcal{S}$
can be expressed by the DC voltage of the voltmeter $U_\mathrm{N}$.
In this formula the device constant of the noise meter $\mathcal{A}$
can be expressed by the pass-bandwidth $\mathcal{B}$.
The pass-bandwidth can be expressed 
from \Eqref{pass_band_width}
by the maximal amplification $Y$ from \Eqref{Y} 
and the bandwidth of the filter.
For educational illustrations the small correction $\varepsilon$
defined by \Eqref{eq:err} can be neglected.
Its evaluation requires numerical integration \Eqref{eq:U2}.
Performing this chain of substitutions
we finally obtain~\cite{epo6}
\begin{align}
q_e=\frac{\md i_\mathrm{N,+}^2}{2\,\md I_\mathrm{ph}}
=\frac{Y_\mathrm{NIA}\mathcal{A}}{2R_\mathrm{N}}\,
\frac{\md U_\mathrm{N}}
{\md U_\mathrm{ph}}
=\frac{2r_\mathrm{G}}{2R_\mathrm{F}+r_\mathrm{G}}
\left(\frac{R_\mathrm{G}}{R_\mathrm{F}^\prime}\right)^{\!\!4}
\frac{R_\mathrm{L}\,C_\mathrm{L}U_\mathrm{m}}
       {(1-\varepsilon)R_\mathrm{N}}
\frac{R_1}{R_1+R_2}
\frac{R_\mathrm{V}+R_\mathrm{av}}{R_\mathrm{V}}
\frac{\md U_\mathrm{N}}
{\md U_\mathrm{ph}},
\label{q_e_final}
\end{align}
cf. Ref.~\cite[Eq.~17]{epo6c}.
The photo voltage $U_\mathrm{ph}$ has to be measured by a voltmeter switched after the buffer before the first capacitors in \Fref{Fig:PreAmp}.
Analogously to \Eqref{Boltzmann_constant}
derivative means the slope of the linear regression in the plot
$U_\mathrm{N}$ versus $U_\mathrm{ph}$.
In such a way the electron charge can be expressed by the 
parameters of the circuit and two voltages.
The experiment consists of the linear regression in the plot
$U_\mathrm{N}$ versus $U_\mathrm{ph}$ when the intensity of light
is changed~\cite{epo6:a,epo6}.
Analogously to this formula for the electron charge we can rewrite 
the derived formula for the Boltzmann constant
\Eqref{Boltzmann_constant} as~\cite{epo5,epo5:a}
\begin{align}
k_{_\mathrm{B}}
=\frac{\mathcal{A}}{4T^\prime}
\left.\frac{\md U_\mathrm{N}}
{\md R_\mathrm{N}}\right|_{R_\mathrm{N}=0}
=\frac{1}{T^\prime}
\left(\frac{r_\mathrm{G}}{2R_\mathrm{F}\,+\,r_\mathrm{G}}\right)
^{\!\! 2}
\left(\frac{R_\mathrm{G}}{R_\mathrm{F}^\prime}\right)^{\!\!4}
\frac{R_\mathrm{L}\,C_\mathrm{L}U_\mathrm{m}}
       {(1-\varepsilon)}
\frac{R_1}{R_1\,+\,R_2}
\frac{R_\mathrm{V}\,+\,R_\mathrm{av}}{R_\mathrm{V}}
\left.\frac{\md U_\mathrm{N}}
{\md R_\mathrm{N}}\right|_{R_\mathrm{N}=0},
\label{k_B_R}
\end{align}
cf. Ref.~\cite[Eq.~17]{epo5},
where we suppose that absolute temperature is already determined
in the used units.

Starting from the fundamental physics and state of the art electronics 
we give comprehensive description of the set-up and the the theory of its work.
This description is not addressed to the students but we give it only because it is new 
and cannot be fuond in the literature. 

Let us make a comparison. Every electronic calculator is a sophisticated device.
The description of all its element is understandable to professionals
but not for the students which take the calculator only to calculate.
The same is for our set-up for determination of the absolute zero.
Imagine it is given to a high-school student.
What should be the instruction which we have to give in order
after some measurements of temperature and voltage by
multimeter students to be able to determine the absolute temperature zero.


\section{Determination of the absolute zero}
    \label{Experiment}
    
Imagine that on the table we have several vessels with water in different temperatures.
It they are of order of 1 liter the change of the temperature will be negligible during the time of the measurements.
In the left, for example, we can put cold water with flowing ice $t^\prime=0^{\circ}$C, 
and on the right a thermos with almost boiling water  $t^\prime=90^{\circ}$C. 
Between 1-3 vessels which interpolate different temperatures in the tea cup interval.
On the table you have to have a towel as well.

Let us switch the first multi-meter as a thermometer and put the thermocouple sensor in the cold water.
Put 4 batteries at the clips and carefully switch them on the set-up.
Be careful not to inverse the polarity because the OpAmps will be burnt.
Then switch on the second multi-meter as a voltmeter no measure to measure 
voltage output.

Immerse the resistor at the end of the triaxial cable in the same water cup.
First measurement is using the short circuit and to write 
the voltmeter indication $U_0$.
For brevity in this section we omit index V in the notation of the voltage
measured by the DC voltmeter in \Fref{Fig:Schema}.
Switch the voltmeter in 200~mV range and work with mV.
Then remove the short-circuit and measure again the voltage $U$
which is now influenced by the thermal voltage noise.
The difference $\Delta U_\mathrm{N} \equiv U-U_0$ 
can be ascribed to the thermal noise 
and it is proportional to the absolute temperature
\be
\Delta U=\mathcal {D}T^\prime,\qquad T^\prime=t^\prime-t_0^\prime,\qquad
\mathcal{D}\equiv\frac{4R_\mathrm{N}\kb}{\mathcal{A}}
=\kb 
\left(\frac{2R_\mathrm{F}+r_\mathrm{G}}{r_\mathrm{G}}\right)
^{\!\!2}
\left(\frac{R_\mathrm{F}^\prime}{R_\mathrm{G}}\right)^{\!\!4}
\frac{(1-\varepsilon)R_\mathrm{N}}
       {R_\mathrm{L}\,C_\mathrm{L}U_\mathrm{m}}
\frac{R_1+R_2}{R_1}
\frac{R_\mathrm{V}}{R_\mathrm{V}+R_\mathrm{av}}.
\label{proportionality}
\ee
Here the index N of the voltages $U_\mathrm{N}$ 
and $U_\mathrm{N,0}$ is for brevity omitted.
Introducing the reciprocal value $\mathcal{C}\equiv1/\mathcal{D}$
we can express the absolute temperature by the measurable 
voltage difference $T^\prime=\mathcal{C} \Delta U$.
The offset of the multiplier is canceled by the described procedure of subtraction from the measured signal the voltage at short circuit input.
For the determination of the absolute zero the calculation of the 
coefficient of the proportionality $\mathcal{D}$ is not necessary.
However, determining the slope of this proportionality after a linear regression,
we can determine the Boltzmann constant at calculated parameter $\mathcal{A}$
\be
\mathcal{D}\equiv\frac{\md (U-U_0)}{\md t^\prime},\qquad
\kb =\frac{\mathcal{A}}{4R_\mathcal{N}}\,\mathcal{D}
=\frac1{R_\mathrm{N}}
\left(\frac{r_\mathrm{G}}{2R_\mathrm{F}+r_\mathrm{G}}\right)
^{\!\!2}
\left(\frac{R_\mathrm{G}}{R_\mathrm{F}^\prime}\right)^{\!\!4}
\frac{R_\mathrm{L}\,C_\mathrm{L}U_\mathrm{m}}
       {(1-\varepsilon)R_\mathrm{N}}
\frac{R_1}{R_1+R_2}
\frac{R_\mathrm{V}+R_\mathrm{av}}{R_\mathrm{V}}\,
\frac{\md (U-U_0)}{\md t^\prime},
\label{k_B_T}
\ee
this is a new result of the present study.
We use again the ``$\equiv$'' sign because the
\Eqref{k_B_T} is the experimental definition of $\mathcal{D}$
as a slope of a linear regression
while \Eqref{proportionality} was the theoretical definition for the 
coefficient of the proportionality $\mathcal{D}$.
We have to mention also that $\md t^\prime= \md T^\prime$
the temperature difference of Kelvin and Celsius degrees 
is equal. The difference is only in the initial of the scales.

\textcolor{black}{
It should be mentioned that described method is not the first simple estimate of the determination of the absolute zero temperature.
From a long time simple demonstration is to use the ideal gas relationship
for the pressure $P$ and volume
$P \tilde{V}=\nu \tilde{R} T^\prime$
in many cases~\cite{Dragia:03}, for the definition of the new Kelvin, for example.
However, the purpose of the present study is not to design a thermometer
but to demonstrate thermal fluctuations.
To determine the Boltzmann constant $\kb$ it is necessary to study fluctuations
or to count atoms (not moles $\nu$), i.e. Avogadro number $\tilde{R}/\kb$.
}

In such a way the Boltzmann constant can be determined
by a measurable slope of the linear regression $\Delta U$ versus 
thermometer temperature $t^\prime$ and the parameters of the 
set-up.
We wish to emphasize that all formulas 
(except small correction $\varepsilon$)
can be derived by derived by elementary calculation
accessible for high-school students.
Mathematician David Hilbert used a simple criterion for understanding
a theorem -- you understand the theorem if you consider 
that the proof is trivial.
In this sense the theoretical physics is also trivial.
The theoretician Lev Landau used to say something in the sense
``I am the biggest trivialiser''.
Following this line, our derivation of the proportionality 
coefficient $\mathcal{D}$ from \Eqref{proportionality}
is also trivial. 
It contains several multipliers, for each of them
we have simple algebraic derivation.
The total number of letters necessary for re-derivation
does not exceed the averaged high-school homework on math.
Only a careful exercise on calligraphy.
The corresponding numerical calculations are better to be done with a computer,
\textcolor{black}{the calculated parameters are given in Table~\ref{tbl:pars}.}
\begin{table}[h]
\centering
	\caption{Table of the calculated parameters of the experimental set-up. }
\begin{tabular}{ c  r }
		\hline
		Calculated or measured parameter & Value  \\ \hline
			$U^*$ & \textcolor{black}{2.50}~V\\
			$\textcolor{black}{\tilde{U}}$ & \textcolor{black}{1.00~V}\\
			$\mathcal{A}$ &  45~aV/Hz \\
			$B$ & 60.72~PHz\\
			$\mathcal{B}$ &55.44~PHz\\		
			$\varepsilon$ & 8.7\%\\
			$\tau_\mathrm{_L} \equiv R_{\mathrm{L}} C_{\mathrm{L}}$ & $4.2 \, \mu$s\\
			$\textcolor{black}{|Y|}$&$1.01\times 10^6$\\
			$\mathcal{C}$ & 2.09~K/mV\\
			$\mathcal{D}$ & 479~$\mu$V/K\\
			$\mathcal{A}/4 (T^\prime=296$~K) & 38~zV/(Hz K)\\
			$Y_\mathrm{NIA}\mathcal{A}/2R_\mathrm{N}$& 22.73~aC\\
			$q_e\mathcal{A}/\hbar$ & 0.068\\
                \textcolor{black}{P, peta} & \textcolor{black}{10$^{15}$} \\
                \textcolor{black}{a, atto} & \textcolor{black}{10$^{-18}$}\\
                \textcolor{black}{z, zepto} & \textcolor{black}{10$^{-21}$}\\
			\hline
\end{tabular}
	\label{tbl:pars}
\end{table}

\begin{figure}[ht]
\centering
\includegraphics[scale=0.9]{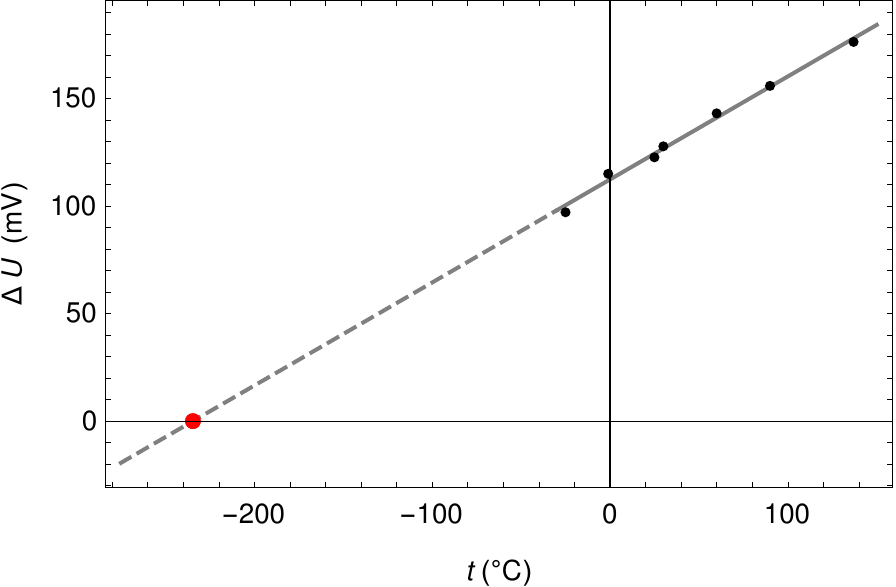}
\caption{The voltage $\Delta U_\mathrm{V}\,$~[mV] measured by voltmeter from
\Fref{Fig:Schema}
versus temperature $t^\prime~[^{\circ}\mathrm{C}]$ of the noise resistor $R_\mathrm{N} = 100~\Omega$.
This is a graphical representation of a table similar to
Table~\ref{tbl:exp}.
The extrapolated interception of $\Delta U=0$ marked by small circle gives estimation $t_0^\prime=-234^\circ$C.
This figure reveals that described set-up measured the spectral density of the thermal voltage noise of a resistor $4R\kb T^\prime.$ The achieved accuracy is of order of
$\epsilon\approx14\%$,
but with a better statistics the accuracy can be increased.
}
\label{Fig:Regr}
\end{figure}

Returning to the determination only the absolute temperature
we describe the first row of Table~\ref{tbl:exp}
\begin{table}[h]
\centering
	\caption{Exemplary table for the measurements.}
\begin{tabular}{ c c c c }
		\hline
		i  & \quad $U$ \quad & \quad $\Delta U$ \quad & \quad $t^\prime$ \quad \\ \hline
		1 &  \dots & \dots & \dots \\
		2 &  \dots & \dots & \dots \\
		\hline
\end{tabular}
	\label{tbl:exp}
\end{table}
Then we can pass to the next vessel and again determine $\Delta U_2$ and $t_2^\prime$.
The purpose of the experiment is to have several points $\Delta U_\mathrm{i},\; t_\mathrm{i}^\prime$,
and to present this table in a plot $\Delta U~$[mV] versus $t^\prime~[^\circ$C].
Draw a line passing near all experimental points in the 
($\Delta U$ versus $t^\prime$) plot.
This procedure is called a linear regression but you can use millimeter paper without knowing the corresponding math.
An example of graphical representation of experimental data is given in
\Fref{Fig:Regr}.

Now you have to make the extrapolation.
This extrapolation represents the accuracy of the measurement using this set-up.
Extrapolating the line far from the experimental points in order to cross the new abscissa
$\Delta U=0$.
Determine at which temperature $t_0^\prime$ is this crossing.
Finally evaluate the accuracy in percents of our measurement
\begin{align}
\epsilon = 100 \times  \frac{\,|t_0^\prime+273|\,}{273}.
\end{align}
In order to demonstrate properties of the setup we make one experiment with laboratory conditions using thermal chamber and good voltmeter with \textcolor{black}{very large}
$R_\mathrm{V}$.
The results are represented in \Fref{Fig:Regr}.
The interception of the line of linear regression 
$\Delta U\propto (t^\prime-t_0^\prime)$ with the abscissa
gives $t_0=-234^\circ$C giving 14\% accuracy which is quite acceptable for 
high-school and undergraduate result which is the purpose of our work,
see for comparison the educational experiment in MIT~\cite{MIT};
cf. also unpublished guides~\cite{MIT1,MIT2,SB}.
\textcolor{black}{By performing} linear regression using a computer,
we easily obtain the uncertainties of the coefficients.
The evaluation of the experimental errors is as a rule tedious for the students,
but our observation is that Jack-knife~\cite{Efron} method they easily accept
and the corresponding programming can be easily performed.

With very careful measurements it is possible to reach percent accuracy,
especially if we include liquid nitrogen to the tea cup temperatures.
Theoretical formulae where only accessories how our set-up completes
its precursors following the story of the physics in the last 115 years.
   
As a whole due to the careful choice of integral circuits,
the accuracy of the described set-up 
is comparable with ones of university physics laboratories.
As additional advance we have given state-of-the-art description of the electronics
at the level of standard university courses and
the budget of remake of the set-up is in pocket range order.

\section{Discussion}
\label{Discussion}

Let us summarize what the ingredients to make this oversimplified experiment which gives satisfactory accuracy and can be performed in high-schools are.
First of all, the appearance of low noise OpAmps with 
$e_\mathrm{N}<1\,\mathrm{nV/\sqrt{Hz}}$
and simultaneously with crossover frequency $f_c>10\,$MHz
gives the opportunity to construct very cheap pre-amplifiers with huge
maximal amplification $Y \gtrsim 10^6$.
Only highly optimized device can reach such amplification
even without screening boxes.
The instrumental amplifier is less sensitive with respect of self-excitation (ringing),
but here we have to emphasize the fundamental importance of the triaxial cable.
The most dangerous parasitic feed-back is the electrostatic coupling of the 
amplified signal to the voltage input of the pre-amplifier.
The triaxial cable substitutes the screening metallic boxes 
and BNC cables which connect the different amplifiers, filters, oscilloscopes.
This simplification gives the possibility the experiment for thermal voltage fluctuations
to gain a virulent mutation in order to
to infect the high schools physics labs from the university ones.
The linear dependence \Eqref{proportionality} and \Fref{Fig:Regr}
describes actually a self-made absolute thermometer
\begin{align}
T^\prime=\mathcal{C}\,\Delta U, \qquad \mathcal{C}=1/\mathcal{D}.
\end{align}
The operational range is limited to the temperatures of use of the triaxial cable
-40$^\circ$C to 200$^\circ$C.
The cable, resistor and soldering of the resistor $R_{\mathrm{N}}$ to cable wires
can survive in immersing in liquid nitrogen,
but when is cooled the cable must not be bent.
Moreover, there is a contemporary tendency the new Kelvin to be
performed by measurements of electric fluctuations and
it is nice and useful high-school students to become familiar with
what actually temperature is and the world in which they will live.
From methodical point of view 
it will be nice 
together with electric demonstration of absolute temperature 
to be performed a demonstration with ideal gasses,
see for example Ref.~\cite{Dragia:03}.
At fixed density $n$ of the particles the pressure $p$ of an
ideal gas  is proportional to the temperature $p=nT$.
This is known for ages.
However, thermal physics is incomprehensible. 
Many physical variables and phenomena, for instance Ref.~\cite{Chatterjee:21}, 
depend on the absolute temperature $T$
and in the simplest possible case of classical statistics the dependence is
just proportionality.
Relatively more recent example, only a half century ago
is the observation by Timko~\cite{Timko:76}
a fundamental property of the silicon transistors 
that: 
``if two identical transistors are operated at a constant ratio of collector current densities, $r$,  then the difference in their base-emitter voltage is $(\kb T/q_e)(\ln r).$ Because both $\kb$ (Boltzman’s constant) and $q_e$ (the charge of an electron) are constant, 
the resulting voltage is directly Proportional To Absolute Temperature (PTAT).''
This article by Timko is the basis for understanding the specification of
a 2-Terminal Integral circuit (IC) Temperature Transducer.
This IC has really excellent properties as an absolute thermometer:
``Linear current output: 1~$\mu$A/K,
wide temperature range: 
-55$^\circ$C to +150$^\circ$C. 
Probe-compatible ceramic sensor package 2-terminal device: 
voltage in/current out Laser trimmed to $\pm 0.5^\circ$C calibration accuracy (AD590M). 
Excellent linearity: 
$\pm 0.3^\circ$C 
over full range (AD590M). 
Wide power supply range: 4~V to 30~V. Sensor isolation from case.''
If the current passes through 1~k$\Omega$ resistor
the calibration gives $\mathcal{D}_\mathrm{AD590}=1\,$mV/K,
a value comparable with our device.
The quoted passages are taken from the AD590 specification~\cite{AD590}.
However, using only an absolute temperature  thermometer it is impossible to determine
neither Boltzmann constant $\kb$ nor electron charge $q_e$
which are byproducts of study of the described noise-meter.
The purpose of our work is not only to design an absolute thermometer but also to demonstrate fluctuations of the electric potential due to
thermal or shot noise.
Roughly speaking, we are making fluctoscopy of the moving electrons analyzing basic notion of the though in thermal physics.

Returning to the evolution, the ideas in physics of Langevin 
random forces are
a convenient tool to study thermal fluctuations.
But when an idea is born we can return to the Democritus 
and its popularization by Lucretius Carus.
The thermal fluctuations of the voltage at the end of a resistor can
be considered as Lucretius Carus \textit{exegium clinamen principiorum}~\cite[Sec.~7.1]{Jammer:67}.

Often a critique can be heard that high-school physics hardly accepts 
new ideas for innovation and modernization.
The world condition in the next years \textcolor{black}{will} not be very favorable 
but at least we can try our best at the present \textcolor{black}{times}.

Perhaps not every physics teacher can make a cover version of the described set-up 
but definitely everyone of them can use it.
Low price gives the unique possibility set-ups to be made in a town 
and send by post to the whole world.
The present article is addressed to the teachers,
but it can be the basis of the guide for the high-school students 
how to measure the absolute temperature zero.
In short, the described set-up is a part of modernization of the
teaching of fundamental physics,
a modernization which became possible by the emergence of low price high-tech 
active electronic components.
Upon request authors can provide the drawings of the PCB,
the list of components and even the purchase orders.

We believe that our simple set-up and accompanying theoretical consideration 
will be interesting and accessible to reproduce by a diverse audience of 
physics students, educators and researchers.
The level of our description is oriented to universities,
but due to simplicity set-up can be reproduced and used even in high schools.
The only device necessary to be added is an affordable multimeter 
which is easily acceptable within pocket money budget. 
And who knows maybe some of the ignited students 
can find career in the physics and even in the metrology.

\section*{Acknowledgments}
The authors are grateful to
Peter~Todorov, Genka~Dinekova, Pancho~Cholakov and Hassan~Chamati, 
for the support and interest in the study, and the creative atmosphere.
This work was supported by the Joint Institute for Nuclear Research, Dubna, RF THEME 01-3-1137-2019/2023 and Grant D01-229/27.10.2021 of
the Ministry of Education and Science of Bulgaria.

\bibliography{zero_temp}

\appendix

\section{Simple derivation for the formulas for amplification}

We use well-known formulas for the amplification of the used basic amplifiers 
but instead to cite textbooks,
we give a simple re-derivation compatible with the used notations following Ref.~\cite{epo5}.
\textcolor{black}{
Formally such material is part of the university education but simple algebra 
can be reproduced by a teenager. 
This supplementary material reveals the close relation between
the basic physics and electronics and we a trying to build a bridge with unique
education between the theoretical analysis of the thermal fluctuations and 
the contemporary electronics. As a rule students of thermal physics are not familiar 
with the necessary for our understanding electronics.
}

\subsection{Basic equation of operational amplifiers (OpAmp)}
We recall the basic equation for the operational amplifiers (OpAmp)
giving the relation between output voltage $U_0$ and the voltages at
plus $U_+$ and minus $U_-$ inputs
\be
U_\mathrm{o}=(U_+-U_-)\,G(\omega),
\qquad |G(\omega)|\gg1.
\ee
\textcolor{black}{In short, an operational amplifier has
two voltage inputs $U_+$ and minus $U_-$ with very small input current.
And one voltage output $U_0$ which can give significant output current current 
so the equation above to be satisfied in linear regime.
In the schematics of the circuits voltage supply terminals for plus $+V_S$ and
minus voltage supply $-V_S$. 
The used by us dual OpAmp ADA4898
has dual ($U_+$, $U_-$, $U_0$) probes
denoted at Fig.~2 of Ref.~\cite{ADA4898}
as +IN1, -IN1, V$_\mathrm{OUT1}$ and +IN2, -IN2, V$_\mathrm{OUT2}$.
}

\textcolor{black}{
The OpAmp consists of sophisticated electronic components: 
transistors, resistors and capacitors hidden in a commercially applicable black box
shown in scale in Fig.~49 Ref.~\cite{ADA4898} as in scale
8-Lead Standard Small Outline Package with Exposed Pad.
In the first approximation Op Amp is just a high gain amplifier
$\alpha=1/G\approx0$ and we have to take into account only the firs real ans imaginary correction.
The implementation of the described set-up become doable after recent 
appearance of cheap, low-noise and high-speed Op Amps.
In such a way this which was doable in Bell laboratories a century ago
now can be in the hand in every student on physics in every school in the world.
Last but not least existence minimum of electronics is important ingredient of teaching of physics. 
Young Richard Feynman used to repair nonworking radios in the beginning.
}

As open loop gain in modulus is a big number is useful to introduce its reciprocal 
value
\be
\alpha(\omega)\equiv\frac1{G(\omega)}\approx\frac1{G_0}+\tau s
=\frac1{G_0}+\mathrm{j}\frac{f}{f_c}
\approx \mathrm{j}\tau\omega,
\qquad
|\alpha(\omega)|\ll1,
\qquad f_c\equiv\frac1{2\pi\tau},\quad \omega=2\pi f,
\ee
which is small in modulus and linear function from frequency $\omega$.
For an order of evaluation one can choose $G_0\approx 10^6$ and $f_c \approx 100$~MHz.
The last approximation is applicable in the interval
\be
\frac{f_c}{G_0}\ll f\ll f_c,
\ee
i.e, in kHz to MHz range which is just our case.
Further we use the basic equation of Op amps in the form~\cite{Ragazzini,master,Manhattan}
\be
\alpha U_\mathrm{o}=U_+-U_-,\qquad 
\alpha\approx\tau s, 
\qquad s=\mathrm{j}\omega,
\qquad \tau=\frac1{2\pi f_c}.
\label{Master}
\ee
Our first problem is the amplification of noninverting amplifier.

\subsection{Non-inverting amplifier (NIA)}

\begin{figure}[ht]
\centering
\includegraphics[scale=0.4]{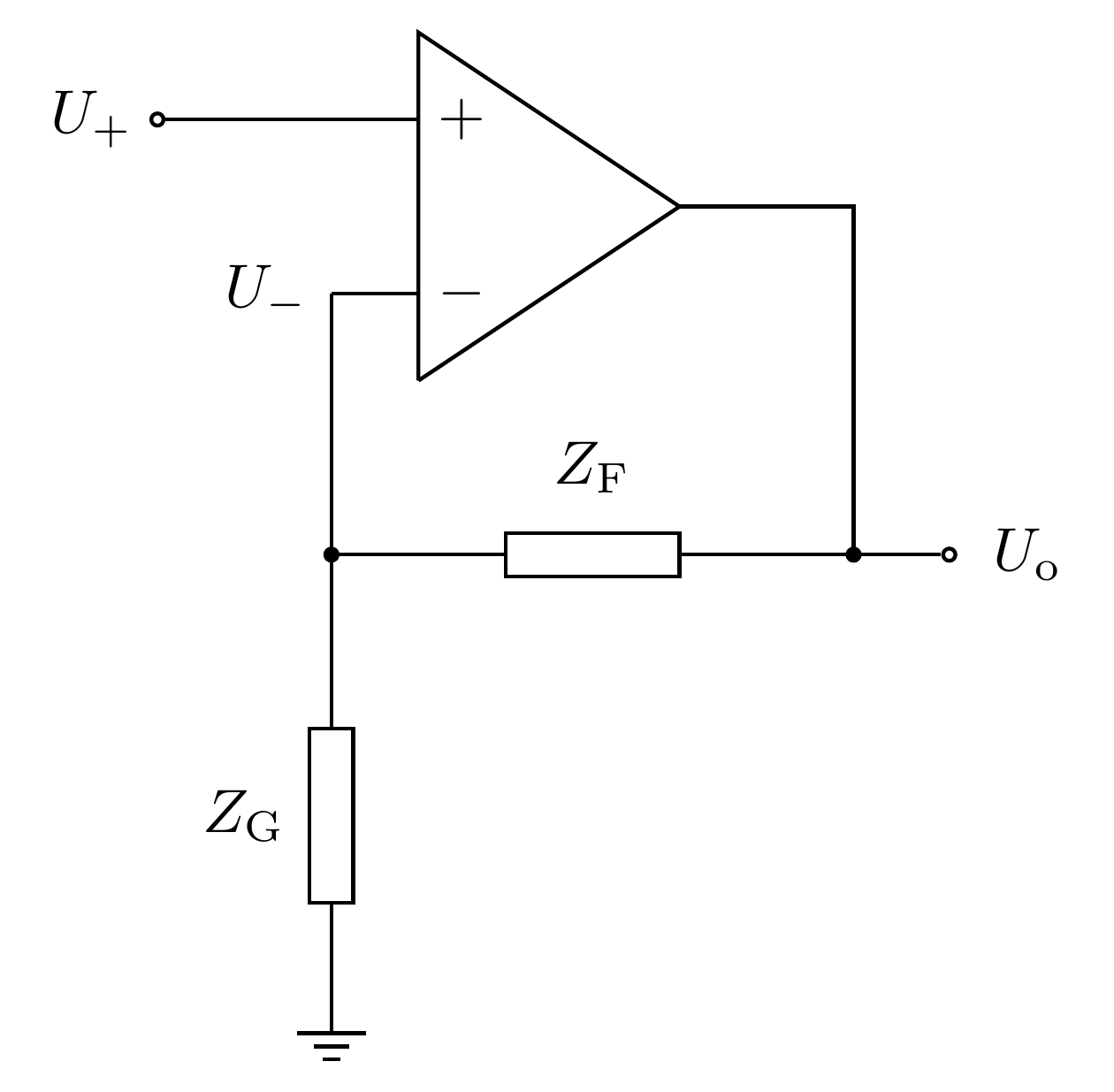}
\caption{High impedance non inverting amplifier $U_\mathrm{i}=U_+$.
The well-known result for the amplification is given by \Eqref{Upsilon_NIA}.
}
\label{Fig:NIA}
\end{figure}

The non-inverting amplifier is drawn in \Fref{Fig:NIA}.
One can see that input voltage of the amplifier is just the plus voltage 
of the OpAmp 
\be
U_+=U_\mathrm{i},\qquad 
U_-=\frac{Z_\mathrm{G}}{Z_\mathrm{G}+Z_\mathrm{F}}\,U_\mathrm{o},
\label{U+-}
\ee
while minus voltage can be expressed as a voltage divider of output voltage
and sequentially connected gain $Z_\mathrm{G}$ 
and feedback $Z_\mathrm{F}$ impedances.
The substitution of \Eqref{U+-} in \Eqref{Master} gives
\be
U_\mathrm{i}-\frac{Z_\mathrm{G}}{Z_\mathrm{G}+Z_\mathrm{F}}\,U_\mathrm{o}
=\alpha\, U_\mathrm{i},
\ee
or
\be
\Upsilon_\mathrm{NIA}\equiv\frac{U_\mathrm{o}}{U_\mathrm{i}}
=\dfrac1{\dfrac1{\dfrac{Z_\mathrm F}{Z_\mathrm G}+1}+\dfrac{s}{2\pi f_c}}
\label{Upsilon_NIA}
\ee
in agreement with the cited well-known formula which we used.
The derivation of the amplification for inverting amplifier (IA)
which we re-derive in the next section also requires elementary high-school algebra which is easy to trace.

The formula for the amplification of the multiplier 
\Eqref{multiplier_amplification} can also be understood considering 
a voltage divider.

\subsection{Inverting amplifier (IA)}

\begin{figure}[ht]
\centering
\includegraphics[scale=0.4]{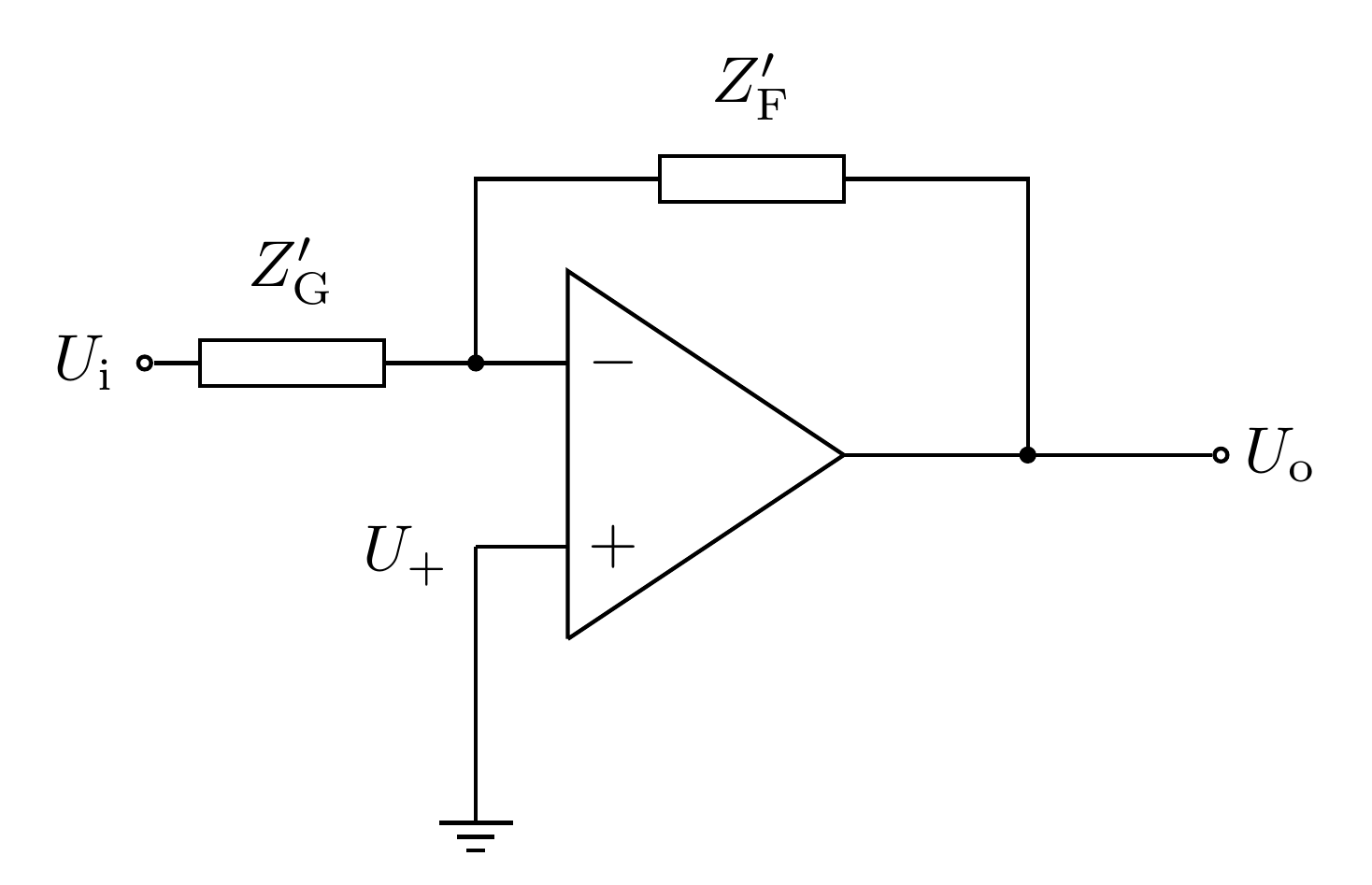}
\caption{Low impedance inverting amplifier with grounded $U_+=0$.
The amplification given by \Eqref{Upsilon_IA} also can be derived by consideration of the elementary algebra of a voltage divider \Eqref{IA_divider}.
}
\label{Fig:IA}
\end{figure}

In the schematics of the Inverting Amplifier (IA) shown in \Fref{Fig:IA}
the plus input voltage is grounded 
\begin{align}
U_+=0,\qquad
U_-=U_\mathbf{o}+(U_\mathrm{i}-U_\mathrm{o})\,
\frac{Z_\mathrm{F}^\prime}{Z_\mathrm{F}^\prime+Z_\mathrm{G}^\prime},
\label{IA_divider}
\end{align}
while for consideration of minus input one can start from output voltage
and to add the difference between the input voltage and output voltage 
which passes through a voltage divider of sequentially connected feedback and 
gain impedances.
The substitution of these formulas for $U_+$ and $U_-$ in the basic equation 
\Eqref{Master} after some elementary calculation gives
the well-known result we have cited
\begin{align}
\Upsilon_\mathrm{IA}\equiv \frac{U_\mathrm{o}}{U_\mathrm{i}}
=-\dfrac1{\dfrac{Z_\mathrm{G}^\prime}{Z_\mathrm{F}^\prime}
+\left(\dfrac{Z_\mathrm{G}^\prime}{Z_\mathrm{F}^\prime}
+1\right)\dfrac{s}{2\pi f_c}}.
\label{Upsilon_IA}
\end{align}

It is not necessary to repeat analogous calculations which gives
almost the same result for the difference amplifier
\be
\Upsilon_\mathrm{\Delta}=-\Upsilon_\mathrm{IA}.
\ee

\subsection{The set-up as lock-in voltmeter}

We derived the formula for the averaged square of the voltage
\Eqref{averaged square} supposing that filtered signal 
is applied to both the input probes of the multiplier.
However if amplified and filtered time dependent signal $U_\mathrm{f}(t)$ is applied 
to the first input probe
\begin{align}
X_1=U_\mathrm{f}(t),\qquad Y_1=U_\mathrm{b}(t)
\end{align}
and to other input is applied some basic time dependent signal 
$U_\mathrm{b}(t)$
for the time averaged signal measured by the voltmeter 
$U_\mathrm{N}$ instead of
\Eqref{averaged square} now reads
\be
\left<U_\mathrm{f}U_\mathrm{b}\right>=U^*U_\mathrm{N}.
\label{lock-in}
\ee
As a rule the basic signal is sinusoidal voltage
\be
U_\mathrm{b}(t)=U_\mathrm{b}\cos(\omega t)
\ee
which creates some subtle response 
\be
U(t)=U_0\cos(\omega t+\theta)=
\Re \left\{U_0\exp [\jm (\omega t+\theta)] \right\}
\ee
in the studied subject.
This signal is then amplified and filtered
\be
U_\mathrm{f}(t)
=U_f \cos(\omega t+\theta+\varphi+\varphi_{_L})
=\Re \left(\Upsilon_\mathrm{LPF} \Upsilon\,U_0
\exp (\jm (\omega t+\theta) )\right), \qquad
U_f=\left|\Upsilon_\mathrm{LPF} \right| \times \left|\Upsilon\right|\times U_0 .
\ee
The substitution of those voltages in the main equation of the
lock-in \Eqref{lock-in} gives
\begin{align}
U_\mathrm{N}=\cos(\varphi+\varphi_{_L}+\varphi_\mathrm{rot}
-\theta)\,
\left|\Upsilon_\mathrm{LPF}\Upsilon\right|\,
\frac{U_\mathrm{b}\,U_0}{2\,U^*},
\end{align}
where $\varphi_\mathrm{rot}$ is and additional tunable phase shift
applied before basis signal to be applied to the lock-in.
Let us take an example.
For $f=\omega/2\pi=10\,$kHz we have
$|\Upsilon_\mathrm{LPF}|\approx 1$, $|\Upsilon|\approx Y=10^{6}$.
Additionally for $R_\mathrm{V}\gg R_\mathrm{av}$ 
the multiplication voltage constant is $U^*=1$,
and with tunable phase shifter we can get $\cos \approx 1$.
In this case let us apply maximal possible for the AD633 multiplier
voltage $U_\mathrm{b}=10\,$V
and to analize can we measure a nanovolt signal of $U=10\,$nV.
The estimation obtained from the above formula gives
additional output voltage $\Delta U_\mathrm{N}\approx 50\,$mV.
This is comparable with the own noise of the device
measures at short circuit input $U_\mathrm{N}\approx 80\,$mV.
In such a way if we measure for one minute 
the output voltages obtained with switched 
on and off small signals the evaluated voltage difference can
be measured by an affordable multi-meter.
Actually even 5~mV one minute modulation can be also detected
and in this sense the considered device
is actually an affordable lock-in voltmeter which every body can do himself
to measure nV signals.
The sensitivity can be significantly improved
by replacement of the low pass filter with a resonance one.
The simplest idea is to substitute the $C_\mathrm{L}$
with a frequency dependent negative resistor (D-element)~
\cite{Bruton:69,Bruton:70,TCAS},
i.e. $1/(\jm\omega C_\mathrm{L})\rightarrow -1/(\omega^2 D)$
and simultaneously to add for stability against ringing a big capacitor
$R_\mathrm{L}\rightarrow 
R_\mathrm{L} +1/(\jm\omega C_\mathrm{G})$.
In the next subsection we will try to analyze some typical
obstacles which prevents progress in the basic physics education.

\subsection{Apology of the formulas}

Let us emphasize again that noise thermometry, and determination
of $\kb$ and $q_e$ are included in the university study education 
for decades.
What is obligatory to all those experimental set-ups.
First of all a good low-noise pre-amplifier.
Many universities can allow to buy it without troubles.
The problem is actually not completely financial.
The commercial amplifier has a clearly written manual 
an within one our every student can use it.

What is our alternative -- to make oneself the amplifier
which requires one day soldering of electronic components
on a printable circuit board.
The corresponding theory is elementary and the total 
derivation does not exceed the volume of a college 
homework on algebra.
The first crack is however obvious
for many students and even for teachers
electronics is a handicraft and on the other side
the derivation of formulas is a tedious math 
similar to those required to pass 
corresponding exams to obtain diploma.

Actually some electronics is obligatory for all physics students
but even in the textbook on the ``art of electronics'' one can
read jocks like ``the generators amplify and the amplifiers generate''.
In any cane a commercial amplifier is a box for the students 
which is opened only in the services.
Our device is actually an open circuit at the screening is
reduced only to the third most external shield of the triaxial 
cable.

In short our purpose is provoke teachers and student to design
his amplifier, to make necessary calculations and to perform 
the assembling. 
In short our purpose is to stimulate creativity.

Let us continue further.
From the amplified signal $U_\mathrm{a}(t)$ it is necessary to evaluate 
the mean square in some time period.
The simplest way is to record the signal by a digital oscilloscope
and further to calculate the corresponding meas values and fluctuations 
by a computer.
No doubts physics students has to be able to use oscilloscopes and computers. 
What is the purpose of our article to stimulate student to design 
something which is working passing several times the non-existing boundary between physics an engineering. 
But for the derivation of the corresponding formulas is necessary
to analyze how the numerical multiplication
can be performed analogously by an integral circuit of analogous multiplier. 
The multiplication squaring and proportion are ingredients
even in the initial school education. 
However the chain of elementary formulas looks for
many students already as a tedious collection.
A motivated student or teacher has to pass and assimilate some 
new notions.
Perhaps the simplest one is the amplification $\Upsilon$ 
of the amplifier followed by a filter which can be frequency dependent. 
The next already new notion is the pass-bandwidth $\mathcal{B}$.
With the multiplication and analogous squaring of the signal
is related the voltage multiplication constant of the device $U^*$.

Most aggressive resistance in our explanation meets the spectral density
of the voltage $\mathcal{S}$ having dimension $\mathrm{V^2/Hz}$.
Most of the university teachers consider that this is notion
belonging of advanced electronics.
This is because the change of the terminology.
Spectral density in sense of intensity per unit frequency
is accepted in the physics education since more that a century
with explanation ob the black-body radiation.
We can refer to Rayleigh, Jeans, Wien, Planck and Einstein
and even before implicitly this notion was used by Hershel 
discovering infra red light.
In conclusion nothing new if we use square voltages per unit frequency.

A motivated young man can use even new for him notions.
In our case they are calibration constants for the noise-meter 
$\mathcal{A}$, 
and the thermometer $\mathcal{C}$ or $\mathcal{D}$.
In order to explain our result and the ability of the described 
set-up we have to use some notions and notations.
In such a way we have addressed to a a little bit philosophical 
problem: 
Whether accumulation of a many trivial ingredients lead
to and non-trivilal result and what is the crucial dose?

The people involved with administration of education 
are as a rule against every innovation.
In some sense their duties are to conserve the status-quo 
and to stop the development.
The places where music is learn are called Conservatory
and as a jocular etymology the mission of the ministry
of science is to conserve the present level of education
and even to worsen it using the polite-correct word 
to alleviate it.
Even the youngest coauthors of the present article 
express disagreement: 
Why the final result for the Boltzmann constant $\kb$ 
\Eqref{k_B_T} and \Eqref{k_B_R}
and electron charge $q_e$ \Eqref{q_e_final} are
written in a doubled manner with short and long notations.
The answer is related with the double purpose
of the article.
On one side we have tho express the final fundamental results
using universal notions as amplification, bandwidth, or constant relating
spectral density of the voltage fluctuations with the measured
by voltmeter DC voltage.
Those notions or their equivalents or synonyms  indispensable 
will be used in the other implementation of the same idea
to organize university lab-work to measure $T$, $\kb$ and $q_e$.
On the other side for detailed description of every set-up 
it is obligatory the final result to be expressed by the parameters 
of the used components and the measurable physical variables.
The relation between these two final goals is one of the main purposes
of our methodological study.

Now let us consider what total dose of efforts is acceptable 
for a self-made set-up for a student lab.
We made many experiments with smart white mice.
The prototypes of the described set-up was used for 
illustrative experiments to the courses of statistical physics for more
then then years.
It was an alternative method students to pass the exam
-- to made from scratch the set-up for determination of 
a fundamental constant  $\kb$ and $q_e$.
It was an advantage that they used most recent integral circuits
which was not used at those time in the commercial electronic devices
as amplifiers and filters.

Developing some modification of the set-ups we attracted as 
co-collaborators even high-school students and 
as a medical experiment we can confirm that one week honest work
is not a lethal dose for a motivated student.
A a by product our co-collaborators obtained additional initial 
speed in their professional career.

Performing experiments with hundreds high-school student we
observed that most of them reading the tutorial to the set-up can
further to perform the experiment and to determine
$T$, $\kb$ and $q_e$.
It was necessary to switch the batteries, integral circuits and
changing resistors or light intensity to measure some voltages by
DC voltmeters used in their high-schools.
Some person of them was able even to reproduce the derivation
of the used formulas and to to understand completely the
used electronics.
The youngest one was able to make some simple test measurements
with the set-up which was also a significant success.

Concerning the eldest potential readers of the present article
which are in the position to design new set-ups for the physics education
their peers are climbing to Mount Everest and even K2 
and with this comparison to make the described set-up 
in one day is like to make a safety weekend trip to 
Mont Blanc.

Finally looking only to the formulas which have to be re-derived 
in order to have comprehensive understanding of the described set-up
in the level to design much better device is necessary to
use several pieces A4 paper and one your.
If we compare theoretical description of our
set-up with complete description
of the electronics theory of a Butterworth filter, commercial amplifier
and a oscilloscope the criterion for simplicity will be obvious.
It is very instructive for the students to understand and feel  
all details of the set-up. 
For the beginning is even better to use self-made ugly
set-up instead a good looking device.
Professional use later will not escape. 

Of course the present manuscript is 
not in the style of the papers written for carrier, 
but it is matter already to a social discussion

\section{Bilinearity of the thermal noise $(\mathcal{E}^2)_f=4 R_\mathcal{N}T$}
\label{Bilinearity}

\begin{figure}[ht]
\centering
\includegraphics[scale=0.9]{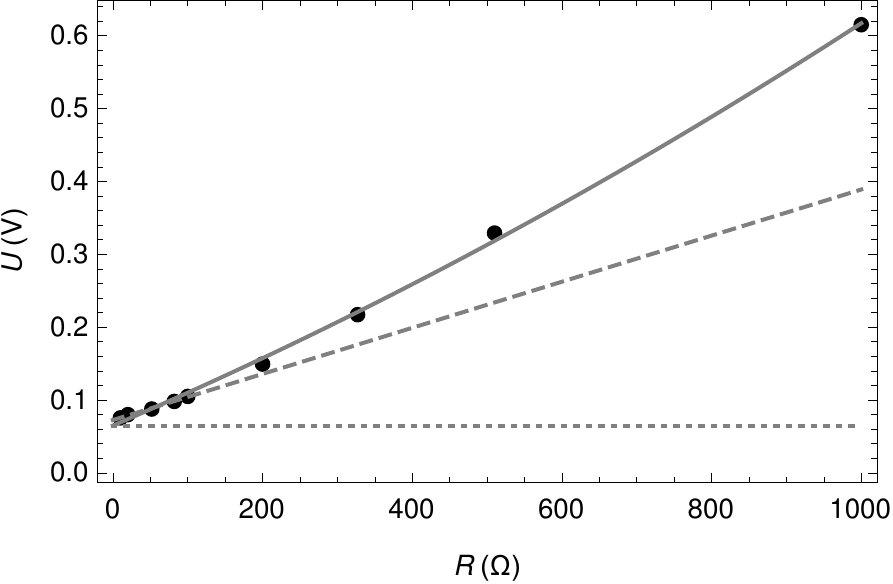}
\caption{Resistance $R_\mathrm{N}$ dependence of 
of the voltage output $U_\mathrm{N}$ of the noise-meter 
which is proportional to 
the spectral density according 
\Eqref{total_voltage_noise}.
The horizontal doted line describes the voltage noise of the OpAmp $e_\mathrm{N}$,
the dashed line describe the linear extrapolation, its slope represents the 
thermal noise.
The continuous line represents the parabolic fi which takes into account the current noise 
of the used OpAmp ADA4898-2.
The fit with high correlation ($>99$\%) gives $\tilde{e}_\mathrm{N}=2.76~\mathrm{nV/\sqrt{2\,Hz}}$
and $\tilde{i}_\mathrm{N}=3.64~\mathrm{pA/\sqrt{2\,Hz}}$ in excellent correspondence with Ref.~\cite[Table~1]{ADA4898}. 
Tilde means our measurement and $\sqrt{2}$ denotes
that two operational amplifiers in the dual OpAmp in the buffer.
}
\label{Fig:Poly}
\end{figure}

The purpose of the present work is to demonstrate determination of the absolute
temperature and to construct an absolute thermometer.
But the linearity of the thermal noise with respect of the resistance is 
an important initial test for the work of the setup.
The thermal noise however cannot be observed separately.
To the measured noise the voltage noise of the first operational amplifier is added
and for big enough noise resistors we have to take into account
the current noise of the OpAmp as well. 
According to \Eqref{V^2/Hz} and \Eqref{spectral_density} referred to input is given by the quadratic polynomial
\begin{align}
U_\mathrm{N} =\frac{\mathcal{B}\,\mathcal{S}}{U^*}
=\frac{R_\mathrm{V}}{(R_\mathrm{V}+R_\mathrm{av})}
\frac{(R_1+R_2)}{R_1U_\mathrm{m}}\,\mathcal{B}\,\mathcal{S},
\qquad
\mathcal{S}=e_\mathrm{N}^2
+4\kb T^\prime R_\mathrm{N}+i_\mathrm{N}^2R_\mathrm{N}^2.
\label{total_voltage_noise}
\end{align}
Our device is actually a noisemeter of the spectral density, 
$U_\mathrm{N}$ is measured with an ordinary DC voltmeter.
Experimental data is represented in \Fref{Fig:Poly}
together with a polynomial fit and a linear extrapolation of the first 5 points,
up to $R = 1~\mathrm{k}\Omega$ including.
As a byproduct of our consideration of fundamental physics
we obtained a new method for measurement the spectral density
of the voltage $e_\mathrm{N}^2$ and current noise $i_\mathrm{N}^2$
of operational amplifiers.
Addressing to the problem to make an absolute thermometer, 
we had to develop a new method for determination 
of current $i_\mathrm{N}$ and voltage $e_\mathrm{N}$ 
noise of the operational amplifiers.
Our parameters agree with the numerical values given in the specification
of the used op amp~\cite[Table~1]{ADA4898}
and in this sense our work can be considered as a
research work in electronics
which is however far beyond our methodical purposes.
We give this appendix only to alleviate the work of colleagues 
constructing similar noise meters.

The analysis of \Fref{Fig:Poly} reveals that it is ignorant to use
noise resistors bigger than 100$~\Omega$.
The instruction of the choice of resistors and OpAmps are described
in Ref.~\cite{LowNoise} for instance.
Actually this is well described by the manufacturers Analog Devices, Texas Instruments and Linear Technology.
Op amps with low voltage noise have to be used only with small enough resistors.
The voltage noise of ADA4898-2 corresponds to thermal noise of say 50$~\Omega$
resistor and that is why we recommend
maximum 100$~\Omega$ for the described thermometer.
The methodological woks are actually  the
most complicated scientific genre of research inspired student 
problems~\cite{Parks:06}.
Before suggesting some innovation in the physics teaching a lot of engineering and scientific work has to be done.

Actually our set-up is a part of a device for measurement of the Bernoulli effect in superconductors~\cite{mPhysC} but \textit{it is another opera,} as the saying goes in some languages.


\section{\textcolor{black}{Measurements with specialised equipment for education}}

We have also performed measurements with specialised educational equipment from Vernier.
This equipment consists of LabQuest~3 data-collection platform with platinum~100 wide range temperature probe sensor and differential voltage probe sensor with $R_\mathrm{V}=20$~M$\Omega$ therefore $U^*=1.05$~V.
The experiment was performed on a relatively warm February evening with
$R_\mathrm{N}=54~\Omega$ with thermalization both in increasing and decreasing temperature.
The resistor $R_\mathrm{N}$ was put in a cigar tubos together with the temperature sensor.
This tubos was put in a box and after the first measurement at ambient outside temperature,
the box was filled with boiling water and a measurement per 15~s was recorded,
this is  Run~1, see additional data file.
The second set of measurements Run~2 was performed after reach of maximal temperature on cooling back to outside temperature, a measurement per 1~minute.
Occasionally, cold water was added to speed up a little bit the decrease rate to ambient temperature.
The results from both sets of measurements are given in Table~\ref{tbl:Vernier} without taking into account the last measurement in Run~2, which was accidentally made upon switching to short circuit for which $U_0 = 130$~mV.
\begin{table}[h]
\centering
	\caption{Table of the results from the measurements with the Vernier equipment, data is given as a supplementary file.}
\begin{tabular}{ l  r r }
		\hline
		 Parameter & Run~1 & Run~2  \\ \hline
		  $t^\prime$~[$^\circ$C] & -253.86 & -273.1 \\
		  corr. coeff.~[\%] & 92.0 & 76.97\\
			\hline
\end{tabular}
	\label{tbl:Vernier}
\end{table}

If one looks carefully at the experimental data or reproduce the linear regression,
will notice the ``quantization'' of the experimental data.
This is no surprise since for our set-up $\mathcal{D}=479~\mu$V/K, 
while the best resolution 
at analog to digital conversion of the voltage probe sensor is 1.6~mV.
This voltage step gives for the temperature quantum
$\Delta t^\prime=3.34$~K,
revealing the meaning of the introduced parameters $\mathcal{C}$ and $\mathcal{D}$.
They show the requirements for the temperature and voltage measurement apparatus for our experimental set-up.
If at the output of the set-up a DC amplifier with amplification
$\times 10$ is added,
the temperature step will decrease to the more precise value of $0.34$~K.
Another solution is to extend the temperature interval between each successive measurement.
Every technical device can be successfully upgraded.

\clearpage

\input{EPO9_en_app.tex}

\end{document}

%% file: EPO9_en_app.tex
\section{Problem of the 9-th Experimental Physics Olympiad, Skopje, 8 May 2022\\
Kelvin's Day}
%
%
\begin{center}
Todor~M.~Mishonov, Emil~G.~Petkov, Aleksander~P. ~Petkov, Albert~M.~Varonov
\footnote{\href{mailto: epo@bgphysics.eu}{epo@bgphysics.eu}}

\textit{Georgi Nadjakov Institute of Solid State Physics, Bulgarian Academy of Sciences\\
72 Tzarigradsko Chaussee Blvd., BG-1784 Sofia, Bulgaria}

\medskip

Leonora Velkoska, Riste Popeski-Dimovski
\footnote{\href{mailto: ristepd@gmail.com}{ristepd@gmail.com}}

\textit{Institute of Physics, Faculty of Natural Sciences and Mathematics,\\ 
``Ss. Cyril and Methodius'' University, Skopje, R.~N.~Macedonia}
\end{center}

\medskip

\begin{center}
\parbox[c]{15.2cm}{
This is the problem of the 9$^\mathrm{th}$ International Experimental Physics Olympiad (EPO), Kelvin's Day.
The goal of the Olympiad is to measure the absolute zero with the given experimental set-up.
If you have an idea how to do it, do it and send us the result;
skip the reading of the detailed step by step instructions 
with increasing difficulties.
We expect the participants to follow the 
suggested items -- they are instructive for physics education in general.
Only the reading of historical remarks given in the first section can be 
omitted during the Olympiad without loss of generality.
\textit{All participants should try solving as much tasks as they can starting from the first one
without paying attention to the age categories: give your best.
The age categories are for ranking purpose only,
work as far as you can.}}
\end{center}
\medskip

Before to begin we wish to emphasize on of the purposes 
of EPO9, to measure the Boltzmann constant.
After the Olympiad please repeat again the experimental 
tasks and you will see how simple it is to measure a
fundamental constant.
As a byproduct you have learned a lot of things 
related to the fundamental physics.
Now you can read the tasks to the end to get to know them and you can start.

\subsection{Description of the experimental set-up 
and the conditions for online participants}
\begin{figure}[ht]
\includegraphics[scale=0.16]{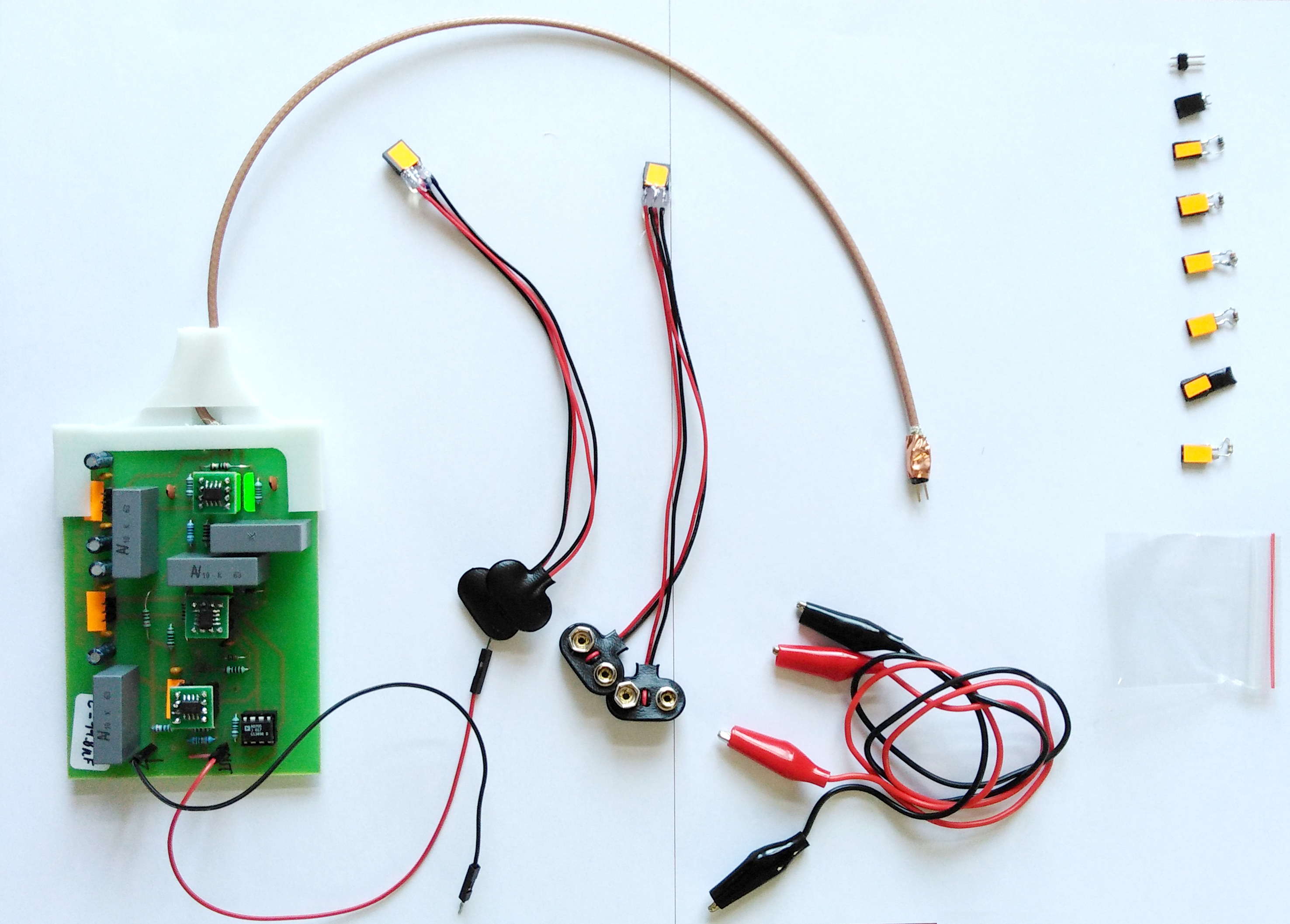}
\caption{Details of the set-up given to the participants:
Printed Circuit Board (PCB) with triaxial cable connected to it (left),
two 9~V battery clip connectors soldered to
3$\times$ PCB female pin headers with orange labels on them (middle),
2 crocodile connecting cables (lower right),
a small plastic bag containing:
one 2$\times$ PCB male pin header (upper right), 
1 short circuit connection on a 2$\times$ PCB female pin headers, and
a set 6 resistors each soldered on 2$\times$ PCB female pin headers,
one of them covered with black electrical tape.}
\label{fig:set-up}
\end{figure}

The set-up you received is represented in the Fig.~\ref{fig:set-up}.
In the next sections different tasks with increasing difficulty 
are described in the sections corresponding to every age category.
The jury will look at the experimental data, tables and graphics.
It is not necessary (nor desirable) to write any humanitarian text between them,
only mark the number of the corresponding task.
We wish you success.

\subsection{Tasks S. Getting to know the resistors set}
\label{Sec:beginning}

\begin{enumerate}

\item 
\label{9V}
Turn on the multimeter as a Direct Current Voltmeter (DCV) 
and measure the voltages of your four 9~V batteries
with maximum accuracy and write them down.

\item 
\label{r}
Turn on the multimeter as an Ohm-meter ($\Omega$).
With maximum accuracy measure the resistance of every of crocodile cables.
Connect them sequentially and measure again its resistance. 

\item 
\label{r}
Again with maximum accuracy measure the resistance of
all given resistors soldered on the pin connectors.
Write on the label the value of each resistor.
If you cannot write with so small numbers just write a character. 
and record the value corresponding of this character.
Order the values of the resistors
in a table like Table~\ref{tbl:R}, 
where i indicates the successive number of the resistor.

We recommend multimeter probes to be directly touched to the resistors.
For the resistor wrapped with adhesive band you can use double pin connector
shown in the upper right corner in Fig.~\ref{fig:set-up}.

If you prefer to use crocodile clips to connect resistors to the probes of the multimeter
you have to subtract 
from the the value given by the multimeter 
the resistance of the crocodile cables.
In such a way you will obtain  the corrected
values of the resistors.

\begin{table}[h]
\begin{tabular}{ c  c c }
		\hline
		&  \\ [-1em]
		i  & $R_\mathrm{i}$ \qquad & $U_\mathrm{i}$ \qquad \\ \tableline
			&  \\ [-1em]
		1 & \dots & \dots \\
		\dots & \dots & \dots \\
\tableline
\end{tabular}
	\caption{Table of the measured resistances of the given set of resistors and their measured voltage noise.}
	\label{tbl:R}
\end{table}
The last column you will fill later.

\item 
\label{167mV}
Calculate how many mV is the difference between $\frac12$~V and $\frac13$~V?
Write the integer rounded value with accuracy of 1~mV.

\item
Using the values in Table~\ref{tbl:values} calculate and write down the 
values of the expressions
\be
Y_\mathrm{NIA}=\frac{R_\mathrm{F}}{r_\mathrm{G}/2}+1,
\qquad
Y_\Delta=-Y_\mathrm{IA}=\frac{R_\mathrm{F}^\prime}{R_\mathrm{G}}, 
\qquad
Y=Y_\mathrm{NIA}Y_\Delta Y_\mathrm{IA}.
\label{Y}
\ee
Those $Y$-values are actually the amplifications of different
segments of the set-up amplifier,
we analyze the theory of its work
as an electronic device.

\begin{table}[h]
\begin{tabular}{ c  r }
		\hline
		&  \\ [-1em]
		Circuit element  & Value  \\ \tableline
			&  \\ [-1em]
			$r_\mathrm{_G}$ & 20~$\Omega$ \\
			$R_\mathrm{F}$ &  1~k$\Omega$  \\
			$C_\mathrm{F}$ &  10~pF  \\ 
			$C_\mathrm{G}$ & 10~$\mu$F \\
			$R_\mathrm{G}$ &  100~$\Omega$  \\ 
			$R_\mathrm{F}^\prime$ & 10~k$\Omega$ \\
			$C_\mathrm{F}^\prime$ & 10~pF \\
			$R_\mathrm{L}$ &100~$\Omega$ \\
			$R_1$ &  2~k$\Omega$  \\ 
			$R_2$ & 18~k$\Omega$  \\
			$R_\mathrm{av}$ & 1.5~M$\Omega$ \\
			$C_\mathrm{av}$ & 10~$\mu$F \\
\tableline
\end{tabular}
	\caption{Table of the numerical values of the circuit elements of the experimental set-up.}
	\label{tbl:values}
\end{table}

\subsection{Tasks M. Determination of the Boltzmann constant}
\item
On every PCB at one of its corners
there is a label with a preliminary measured capacity $C$ written;
it is close to 43~nF.
Write down the value $C$ for your set-up
and from Fig.~\ref{Fig:Eps} determine $\varepsilon$ for it both in \% and absolute value.
\begin{figure}[h]
\centering
\includegraphics[scale=0.8]{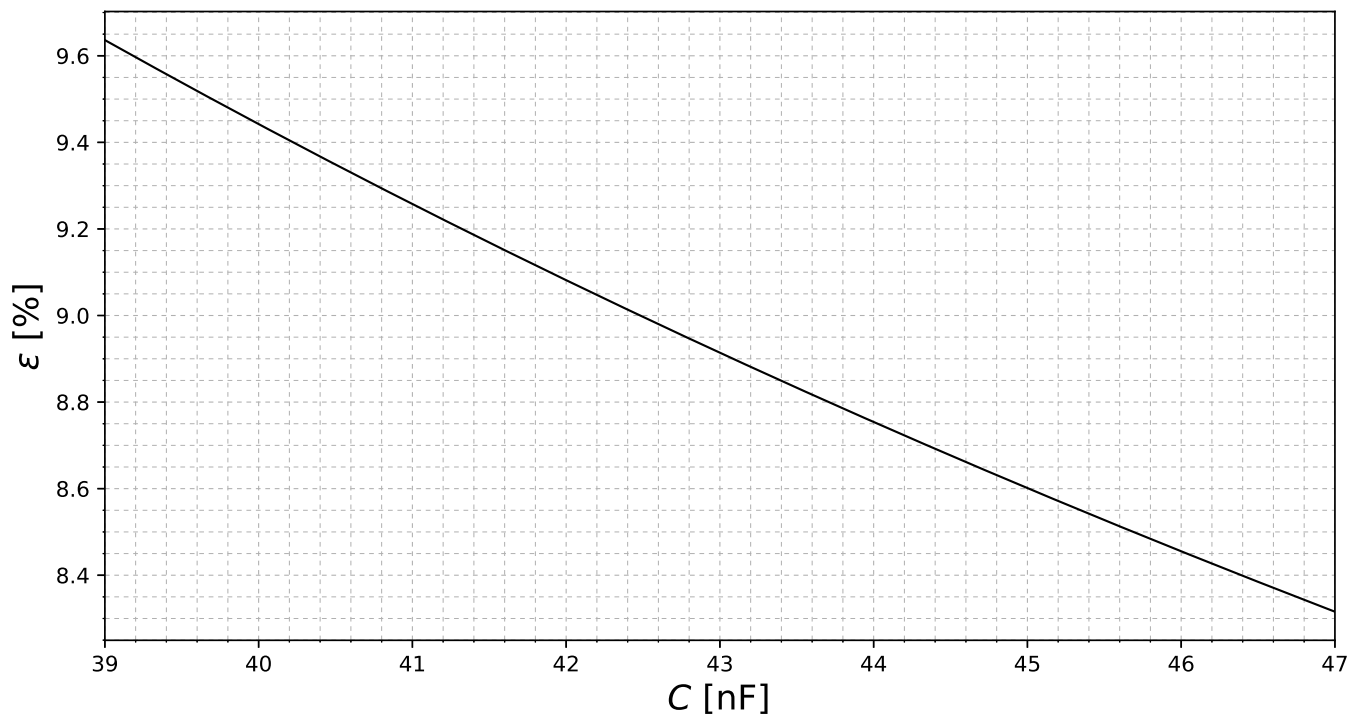}
\caption{Error $\varepsilon$ in percent in determination of $\kb$ as a function of the  
capacitor $C$.}
\label{Fig:Eps}
\end{figure}
Then calculate $(1-\varepsilon)$ for your set-up.

\item
Take the value $R_\mathrm{L}$ from Table~\ref{tbl:values}
and calculate the quantity 
\be
\tau_\mathrm{L}=R_\mathrm{L}C.
\ee
What is the dimensionality of this parameter?
You can use standard mili (m) and micro ($\mu$) which are standard 
for the used SI units.

\item 
Calculate and write down the value 
\be
B = \frac{|Y|^2}{4\tau_\mathrm{L}}
=\frac{|Y|^2}{4R_\mathrm{L}C}.
\label{B}
\ee
What is the dimensionality of this parameter?
This is a very big number,
you can rewrite your result using standard SI notations:
k=kilo=10$^3$, M=mega=10$^6$, G=giga=10$^9$, T=tera=10$^{12}$, P=peta=10$^{15}$, 
E=exa=10$^{18}$.

\item
Now calculate and write down the corrected value
\be
\mathcal{B}=(1-\varepsilon)B
\ee
in PHz. 
In electronics the parameter $\mathcal{B}$ is known as bandwidth 
from where the notation $B$ is used.
Knowing its value is crucial for our experimental task to measure the Boltzmann constant.
Now we can go to our final task.

\item
Switch the multimeter as voltmeter and using one crocodile connecting cable
connect the common points of the voltmeter (usually marked as COM) 
with the cable from the PCB denoted with the
usual triangular sign for ground \tikz{\node[sground,scale=0.5]{}}.

\item
Analogously, connect the other crocodile cable to the voltage output cable
marked on the PCB with ``OUT'' to the voltage input of the multimeter
often marked as V$\Omega$mA.
You will work in the mV range of your multimeter,
if you measure voltages larger than 200~mV, there is a problem
either in connectivity or in the PCB itself.
In such a case, before asking for assistance,
check that the 2 female pins of the triaxial cable are connected by either a resistor or the short circuit.

\item 
Take the resistor with biggest resistance (and biggest number in your table) 
and connect it at the end to the of the three-axial cable 
of the set-up.
This is the voltage input of the set-up.
We actually measure the random voltage which thermal fluctuations induce
at the end of the resistor.
This random voltage is analogous to the Brownian motion of small particles
immersed in a liquid.
If our hearing were infinitely sharp, we would hear the drumming of air molecules on our eardrums.
And today we measure the random motion and voltage of electrons moving along the length of a resistor.

\item Connect the four (2 by 2) 9~V batteries to the 2 doubled connection clips by buckling the electrodes.

\item 
\textbf{Be careful! 
From this point on you can burn the operational amplifiers which will terminate your experiment.}
\textbf{The \underline{orange} labels of the batteries voltage clip connectors
should match with the \underline{orange} labels of the two 3$\times$ PCB male pin headers 
on the PCB these are voltage pins.}
Insert the 2 battery voltage clips in the 2 voltage pins,
the orange labels on the clips \textbf{must} face the orange labels of the pins.
Wait 5 minutes and measure the output voltage.
Make 5 records with interval 1 minute.
Your experimental data should be rewritten and 
ordered in a table 
similar to example Table~\ref{tbl:R-U}.
\begin{table}[h]
\begin{tabular}{ c c c c c c}
		\hline
		&  \\ [-1em]
		i  
		& $U_\mathrm{a}$ \qquad 
		& $U_\mathrm{b}$ \qquad 
		& $U_\mathrm{c}$ \qquad 
		& $U_\mathrm{d}$ \qquad 
		& $U_\mathrm{e}$
		\\ \tableline
			&  \\ [-1em]
		0 & \dots & \dots & \dots & \dots & \dots\\
		1 & \dots & \dots & \dots & \dots & \dots\\
		\dots & \dots & \dots & \dots & \dots & \dots\\
\tableline
\end{tabular}
	\caption{Table of the measured voltages
	for every numbered resistor. 
	For every resistor the median voltage 
	should be  underlined.
	Sequence of the measurements is opposite.
	The measurement starts with the resistor with biggest 
	resistance and finishes with the short circuit
	numbered with zero and its median for the 5    	measurements 	is denoted by $U_0$.
	}
	\label{tbl:R-U}
\end{table}

\item
Then change the resistor, wait 2 minutes 
and make again 5 records every 1 minute.
In such a way you will spend 7 minutes for every resistor.
In these records take the median 
(the voltage biggest than 2 smallest
in smaller that 2 biggest voltages); 
underline this value.

\item
The short circuit should be the last
in your measurements.
Again wait 2 minutes, measure the voltage,
wait 1 minute and measure again.
This will take the last 6 minutes of this series of
measurements; 
(7*6=42)+3=45 
and within one hour you can complete
this essential part of the experiment.
The median of the measurements for every 
resistor has to be written as 3$^\textrm{rd}$.

\item
For half an hour, within one hour you will have median averaged voltage for your set of resistors 
including the short circuit which you should denote with $U_0$. 
Write the median voltage as 3$^\mathrm{rd}$ column in your Table~\ref{tbl:R}.

\item
Using the multimeter probes measure the voltage of every 9~V
battery supplying current to the setup.
Disconnect the batteries from the clips and measure their 
voltage again.
Present these 8 voltages in a table;
before and after disconnecting.
The voltage source should be preserved for later measurements.
Then we can start the experimental data processing.

\item
The results for the averaged voltage has to be graphically presented versus the resistance.
The resistance $R_\mathrm{i}$ is in the abscissa the median voltages $U_\mathrm{i}$
on ordinate
index 0 and corresponding voltage $U_0$
is for the short circuit.
You have to use paper with squares, millimeter or even 5~mm from the notebook.
Stop to think for a while, 
the choice of the intervals is most important 
for the accuracy of our further work.
You will have several points on the $U$-$R$ plot.
Using a ruler draw a line which best approximate the experimental points;
this is almost an art.

\item
\label{lin}
Choose 2 points ``A’’ and ``B’’ close to the opposite ends of the drawn line.
Estimate the resistances
$R_\mathrm{A}$ and $R_\mathrm{B}$ (the abscissas) and
the voltage values
$U_\mathrm{A}$ and $U_\mathrm{B}$ (the ordinates).
Let us denote the voltage difference between these points as $\Delta U$
and analogously difference in resistance by $\Delta R$.
Now we can calculate the most important for our measurement parameter
\be
i_\alpha \equiv \frac{\Delta U}{\Delta R}=
 \frac{U_\mathrm{B}-U_\mathrm{A}}{R_\mathrm{B}-R_\mathrm{A}},
\ee
having dimension current V/$\Omega$=A.
This procedure for approximating of a cloud of experimental points by a line is called in mathematics
linear regression.

\item
Measure the room temperature with the thermocouple, calculate its absolute temperature value
and write them down both.
For example if the room temperature is 23 degree
$T^\prime\approx 273+23=296\,$K. 

\item
Take from Table~\ref{tbl:values} the necessary parameters $R_1$ and $R_2$ for the set-up.
and use the value $U_\mathrm{m} = 10~\mathrm{V}$ which describes the work 
of the integral circuit which makes analogous multiplication.
Calculate the Boltzmann constant, 
including its dimension,
using the formula 
\be
\kb
=\frac{i_\alpha U_\mathrm{m}}{4\mathcal{B}T^\prime } 
\frac{R_1}{R_1+R_2}.
\label{kB}
\ee
The detailed derivation of this formula is given
in the scientific description of the used 
set-up in the main article above.
Congratulation it is perhaps your first measurement of a fundamental constant!


\item
Calculate the product $T=\kb T^\prime$ in Joules, or in SI design
in zepto Joules; 1\,zJ=10$^{-21}\,$J.
This very small energy determines the typical energy of thermal motion.
For example averaged translational kinetic energy of one air molecule is
$E_\mathrm{kin}=\frac32 T$.

%
%
%
%

\subsection{Tasks L. Determine the absolute zero}
This is an extremely difficult task, almost an adventure.
We are using a pre-amplifier with a huge $10^6$ voltage 
amplification.
All precise electronics is in screened metal boxes
but our set-up is not.
That is why this experiment could be easily corrupted
by every external source of noise:
working electromotors, contact of room voltage supply,
luminescent lamps etc. 
Try to minimize all suspicions source of noise 
close to your table or desk.

\item
In the given set of the resistors one is with
the electric insulation tape.
Put more insulation tape again if it is necessary.
Switch this resistor at the end of the triaxial cable.

\item
Compare your obtained value of $U_0$ with 
your median voltage for this resistor.
If $U_0$ is larger than the median value,
your measurement of $U_0$ is not good.
In this case, calculate $U_0$ from the linear regression
using the formula
\be
U_0 = \frac12 (U_\mathrm{A}+U_\mathrm{B})-
\frac{i_\mathrm{\alpha}}{2} (R_\mathrm{A}+R_\mathrm{B}).
\label{U0}
\ee
You may perform this task even if your measurement of $U_0$ is good.
Then compare both obtained values of $U_0$ and decide which one to use or use their average.

\item
Using insulation tape 
wrap together the 
insulated resistor with 
the end of the of thermocouple of 
the temperature sensor.
The resistor and the thermocouple have to be close 
each other. 
The thermocouple will measure the temperature
around the resistor.

\item
Using appropriate vessel heat the salt to around 100$^\circ$C.
The temperature should be definitely below 200$^\circ$C.

\item
Take a tea cup and return to the setup.
Now you have to be very careful.
Put the wrapped thermocouple 
in the center of the tea cup
and carefully put the the hot salt 
around the resistor and thermocouple.

\item 
The most important part of the experiment begins.
Each 2$^\circ$C write in a table with 4 columns:
1) number, 2) temperature, 3) the voltage and
4) reduced voltage as shown in Table~\ref{tbl:t-U}.   
\begin{table}[h]
\begin{tabular}{ c c c c }
		\hline
		&  \\ [-1em]
		i  
		& $t_\mathrm{i}$ \qquad 
		& $U_\mathrm{i}^\prime$ \qquad 
		& $\tilde{U}_\mathrm{i} = U_\mathrm{i}^\prime-U_0$ 
		\\ \tableline
			&  \\ [-1em]
		0 & \dots & \dots & \dots \\
		1 & \dots & \dots & \dots \\
		\dots & \dots & \dots & \dots \\
\tableline
\end{tabular}
	\caption{Table of the measured voltages
	as a function of temperature. 
	}
	\label{tbl:t-U}
\end{table}
When the temperature reaches 40$^\circ$C
remove the batteries from the set-up as
the experimental part is completed.

\item
Fill the 4$^\mathrm{th}$ row in the last table.
The measured voltage minus $U_0$.
The voltage $U_0$ is the median of the voltage 
of short circuit. 

\item
Present the experimental data graphically.
In the abscissa (horizontally) the temperature
$t$ in $^\circ$C,
in the ordinate (vertically) the reduced
voltage $\tilde{U}$.

\item
Draw a line close to the experimental points,
perform a linear regression just like in Task~\ref{lin}
using the temperature interval (40$^\circ$C, 80$^\circ$C).
Calculate the slope
\be
K \equiv \frac{\Delta \tilde{U}}{\Delta t}=
 \frac{\tilde{U}_\mathrm{B}-\tilde{U}_\mathrm{A}}{t_\mathrm{B}-t_\mathrm{A}}.
\ee

\item
Now calculate the intercept,
for which the abscissa value $=0$
($\tilde{U}=D+K t$)
\be
D = \frac12 (\tilde{U}_\mathrm{A}+\tilde{U}_\mathrm{B})-
\frac{K}{2} (t_\mathrm{A}+t_\mathrm{B}).
\ee

\item 
Calculate 
\be
t_0 = -\frac{D}{K}
=t_\mathrm{A}
-\frac{\tilde{U}_\mathrm{B}-\tilde{U}_\mathrm{A}}{t_\mathrm{B}-t_\mathrm{A}}\,
\tilde{U}_\mathrm{A}.
\label{t0}
\ee
This is actually your today evaluation 
of the absolute zero.

Do not be disappointed -100$^\circ$C and
-400$^\circ$C are good results for 
your first determination of the absolute zero.

\subsection{Tasks XL. Derivation of the used formulae}

\item
Derive Eq.~(\ref{U0}).

\item
Derive Eq.~(\ref{t0}).

\item
Derive
\be
\tilde{U}(t) = \tilde{U}_\mathrm{A}+
\frac{\tilde{U}_\mathrm{B}-\tilde{U}_\mathrm{A}}
{t_\mathrm{B}-t_\mathrm{A}}(t-t_\mathrm{A}).
\ee

\item
Using the spectral density of the thermal voltage noise
of a resistor $4\kb T^\prime R$ derive the formula for the determination of Boltzmann constant
Eq.~(\ref{kB}).
For tutorial use the main unabridged article above.

\end{enumerate}

\textbf{Underline your results, scan or photo all 
your pages and send the file in PDF format
to EPO@bgphysics.eu.
File name is your name and category you decide to participate, for example, ``EnricoFermiXL.pdf’’.
}\\

We will announce the results tomorrow.\\

With best regards,\\

EPO

\subsection{Epilogue}
EPO9  is held in a even in a more difficult for the whole world conditions than EPO8.
Approximately half of the participants are on-line.
The organizers of EPO9 would like to thank everyone 
who helped in the preparation of this wonderful competition especially
the president of the Society of Physicists of Macedonia, Prof.~Lambe Barandovski
for ensuring on-spot participation in Skopje.
We are waiting you at EPO10; next year same time.


\subsection{Acknowledgments}

The authors are grateful to
to Peter Todorov, Genka Dinekova, Pancho Cholakov and Hassan Chamati, for the support and assistance,
Beca Natan and Nedeltcho Zahariev for the assistance in the preparation of the experimental set-up for the Olympiad,
to Danica Krstovska and Krastyo Buchkov for the interest,
to Lambe Barandovski and Boce Mitrevski for the hospitality and assistance in the holding of the Olympiad.
Last but not least, we appreciate measurement of the fundamental Boltzmann constant and zero temperature alongside the absolute champion by the participants in the Olympiad:
Teona Ana Georgievska, Matea Mitrevska and David Salontaji (RN Macedonia),
Beatrice Cristea (Romania),
Iosif Keventzidis and
Odisseas Kominis Altanis (Greece),
Nikola Stambolic (Serbia).
Their works confirms the success of fundamental physics experiments in high school physics.

%
%
\subsection{EPO -- a historical perspective. Skip this section during the Olympiad}

From its very beginning, the Experimental Physics Olympiad (EPO) is worldwide known;
all Olympiad problems have been published in Internet~\cite{EPO1,EPO2,EPO3,epo4:a,epo5:a,epo6:a,epo7:a,EPO8} 
and from the very beginning there were 120 participants.
In the last years high-school students from more than 10 countries participated and the distance between the most distant cities 
is more than 4~Mm.

Let us describe the main differences between EPO and other similar competitions.
\begin{itemize}
\item Each participant in EPO receives as a gift from the organizers the set-up, 
which one worked with.
In such a way, after the Olympiad has finished, even bad performed participant is able to repeat the experiment and reach the level of the champion.
In this way, the Olympiad directly affects the teaching level in the whole world.
After the end of the school year, the set-up remains in the school, where the participant has studied.
\item Each of the problems is original and is connected to fundamental physics or the  understanding of the operation of a technical patent.
\item The Olympic idea is realized in EPO in its initial from 
and everyone willing to participate from around the world can do that.
There is no limit in the participants number.
On the other hand, the similarity with other Olympiads is that the problems are direct illustration of the study material and alongside with other similar competitions mitigates the secondary education degradation, which is a world tendency.
\item One and the same experimental set-up is given to all participants but the tasks are different for the different age groups, the same as the swimming pool water is equally wet for all age groups in a swimming competition.
\item One of the most important goals of the Olympiad is the student to repeat the experiment at home and to analyze the theory necessary for the understanding.
In this way any even badly performed motivated participant has the possibility to be introduced to the corresponding physics field, even though there is no physics classroom in his/her school, even though the physics education in his/her country to be deliberately destroyed.
\end{itemize}

We will briefly mention the problems of former 7 EPOs: 
1) The setup of EPO1~\cite{EPO1} was actually a student version of the American patent for auto-zero
and chopper stabilized direct current amplifiers.
It was notable that many students were able to understand the operation of an American patent without special preparation~\cite{chopper}.
2) The problem of EPO2~\cite{EPO2} and EPO8~\cite{EPO8} was to measure Planck constant by diffraction of a LED light by a compact disk.
3) A contemporary realization of the assigned to NASA patent for the use of negative impedance converter for generation of voltage oscillations was the set-up of EPO3.~\cite{EPO3}
4) EPO4~\cite{epo4:a} was devoted to the fundamental physics -- to determine the speed of light by measuring electric and magnetic forces.
The innovative element was the application of the catastrophe theory in the analysis of the stability of a pendulum.
5) The topic of the EPO5~\cite{epo5:a} was to measure the Boltzmann constant $\kb$ following the Einstein idea of study thermal fluctuations of electric voltage of a capacitor.
6) The EPO6~\cite{epo6:a} problem can be considered as a continuation of the previous Olympiad.
With a similar electronic circuit Schottky noise is measured and his idea for the determination of the electron charge is realized.
7) The EPO7~\cite{epo7:a} problem was to measure a large inductance made by a general impedance converter by the Maxwell-Wien bridge.
8) The EPO8 problem was a modified repeat of EPO2 with first online participation in the complicated world global situation in 2021.
 
Each problem given at EPO can be considered as a dissertation in methodology of physics education.

In short, the established traditions is a balance between contemporary working technical inventions and fundamental physics.

The EPO problems are meant for high school and university students 
but are posed by teachers with co authorship 
with colleagues working in universities or scientific institutes.
For colleagues interested in new author's problems for the needs of the contemporary physics education we share our experience in the description of the experimental set-ups described at a university level.
These are for instance:

1) The determination of the Planck constant without light but only with electronic processes study;~\cite{EJP_Planck} this set-up requires the usage of an oscilloscope but in some countries the oscilloscopes are available in  in the high schools physics labs and the prices of the former is constantly going down.

2) The speed of the light without the usage of scales or high frequency equipment is another innovative set-up~\cite{epo4} for high school education.
And the idea for this experiment is given by our teacher in electrodynamics Maxwell.

3) In the physics curriculum in all countries it is mentioned that the temperature is a measure of average kinetic energy of the gas molecules but the Boltzmann constant $k_\mathrm{B}$ that gives the relation between energy and temperature is not measured in high school and even rarely in the best universities.
The experimental set-up for $k_\mathrm{B}$ by the method proposed by Einstein (the EPO5 problem)
is described~\cite{epo5} as a set-up for university school lab exercise in a impact factor journal.
But what larger impact an experimental set-up that is used by high school students from Kazakhstan and Macedonia and the surrounding countries can have.
More than 100 set-ups were distributed around the world.

4) Similar thoughts can be expressed for the electron charge $q_e$.
This fundamental constant is also mentioned in the high school education as a humanitarian incantation but is not measured.
We broke this tradition and described an experimental set-up (from EPO6) in the European Journal of Physics~\cite{epo6}.
This set-up can be built for a week in every high school.
The Schottky idea for determination of the electron charge by measuring voltage fluctuations is used.
From the idea to the realization more than 100 years have passed and one of the reasons is that in many countries the largest enemy of the education is the ministry of education.

The development of some set-ups required additional study of circuits with operational amplifiers (OpAmps).
This led to introduction of the master equation of OpAms
applied to the stability of circuits with Negative Impedance Converter
(NIC)~\cite{master},
Generalized Impedance Converter~\cite{GIC}
and study of Probability Distribution Function~\cite{PDF}
of the crossover frequency of OpAmps.
Continuing this stile, the scientific description of the 
set-up of the present Olympiad (EPO9) is given
at university education level~\cite{epo9}.

The mission of the physics teachers in the worldwide progress is evident -- precisely our science reshaped the world in the last century.
Successful innovative EPO set-ups after some update can be manufactured by companies specialized in production of educational equipment like 
TeachSpin\footnote{\url{https://www.teachspin.com/}} and PHYWE\footnote{\url{https://www.phywe.com/}} for instance. 

\clearpage

\section{Solutions to the EPO9 Problem by the absolute champion Kristijan Shishkoski}

\includepdf[pages=-,width=\textwidth,noautoscale,offset=-10 -150]{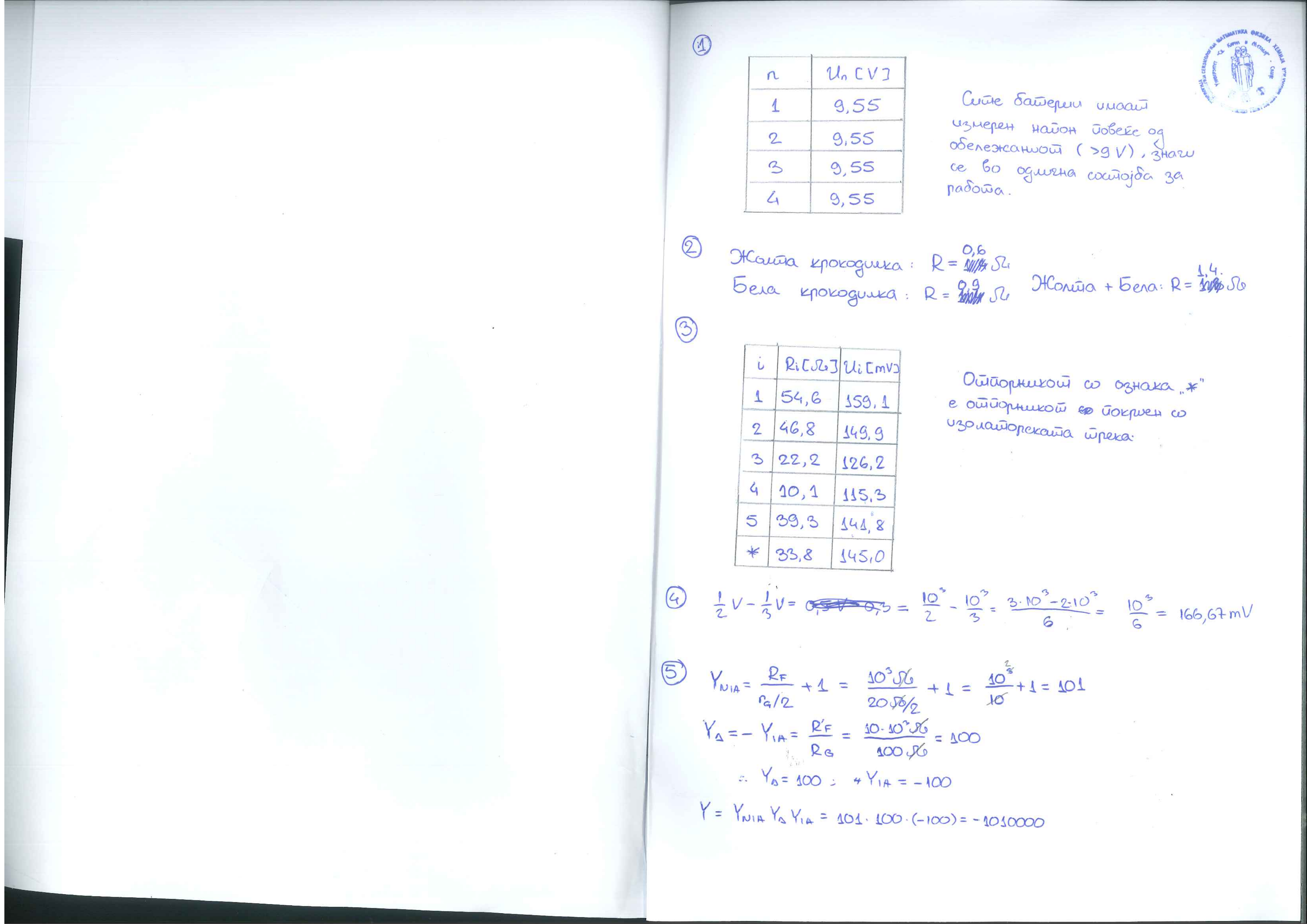}